\documentclass[final,3p,times,twocolumn]{elsarticle}

\usepackage{graphicx}
\usepackage{amssymb}

\usepackage{bm}

\biboptions{sort&compress}

\journal{Special Issue in Physica D}

\usepackage{amsmath, subfigure, hyperref, xcolor}

\begin{document}

\begin{frontmatter}


\title{Arnold web and dynamical tunneling in a four-site Bose-Hubbard model}



\author{Sourav Karmakar, Srihari Keshavamurthy}

\address{Department of Chemistry, Indian Institute of Technology, Kanpur, Uttar Pradesh, 208016 India}

\begin{abstract}
We present the quantum and classical study of a four-well trapped Bose-Einstein condensate (BEC) modeled by the Bose-Hubbard Hamiltonian. The model used is fashioned as a minimal bipartite system consisting of a trimer coupled weakly to a monomer.  Semiclassical insights into the fragmentation dynamics of the condensate is obtained by mapping out the Arnold web of the high dimensional classical phase space.  A remarkable one-to-one correspondence between the quantum state space in occupation number representation and the classical Arnold web in action space is observed. The relevance of dynamical tunneling for the dynamics of Fock states near resonance junctions on the Arnold web is highlighted. We also predict the possibility of manipulating the transport within the trimer subsystem due to the weak coupling to the monomer. 
\end{abstract}


\end{frontmatter}


\section{Introduction}
\label{sect:intro}

Since the experimental realization of Bose-Einstein condensate (BEC) \cite{Anderson198,KetterlePRL95,HuletPRL95}, the stability of a trapped BEC has been studied extensively. The Bose-Hubbard model has been particularly useful\cite{JAKSCH200552,Bloch_NP_2005,Gati_2007} in understanding several aspects of the trapped BECs. Interestingly, insights into the fragmentation of the two\cite{ZiboldPRL2010,OlbrichPRL1996,KellmanPRA2002,RubeniPRA2017,ReinhardtPRA2005,MayaCohenPRA2010} and the three \cite{MunroPRA2000,PennaPRE2003,LiuPRA2007,DeyVardiPRL2018} well trapped BECs have been gained from analyzing  the classical dynamics associated with the Bose-Hubbard Hamiltonian (BHH). As the classical limit Hamiltonian is nonlinear, phenomenon such as bifurcations, chaos leave their imprint on the quantum dynamics of the condensate. For example, in a two well trapped BEC the experimentally observed\cite{ZiboldPRL2010} transition from Rabi to Josephson regime is the quantum manifestation of the classical bifurcation of a fixed point\cite{SmerziPRL1997}. In addition, the emergence\cite{Oberthaler2005} of the macroscopic quantum self trapped states (MQST) can be predicted using the classical Hamiltonian\cite{RaghavanPRA1999}. 

The classical Hamiltonian of a two well BEC has effectively one degree of freedom  and it is classically integrable. However, addition of another well results in a two degree of freedom classical Hamiltonian, which in general is non-integrable and hence capable of exhibiting mixed regular-chaotic dynamics. The emergence of classical chaos in a non-integrable system has its own unique influence on the condensate dynamics. For example, the condensate stability in a three well trapped BEC \cite{MunroPRA2000,PennaPRE2003,LiuPRA2007,DeyVardiPRL2018,kolovsky2007} can be predicted by studying the nature of underlying phase space; thus, the existence of regular regions in the phase space leads to a stable trapped condensate\cite{PennaPRE2003,LiuPRA2007}. Note that the trimeric system leads to several interesting possibilities such as a transistor like behaviour\cite{Stickney_PRA2007}, as an atomtronic switching device\cite{Wilsmann_CommPhys2018}, and as a source-drain transport of single atoms\cite{Schlagheck_2010}. 

Compared to the dimer and trimer,  fewer studies have been made on trapped four well BEC, particularly from a semiclassical perspective. We mention a couple of studies that have investigated the presence of coherence oscillations \cite{OlsenPRA2011,KhripkovPRA2014} and thermalization in model bipartite monomer-trimer systems \cite{KhripkovJPCA2016,KhripkovPRE2018,KhripkovNJP2015}. The paucity of studies on multiple ($>3$) well BEC from a semiclassical perspective is in part due to the fact that such systems result in classical multiresonant Hamiltonians with three or more degrees of freedom. The nature of classical transport in such systems is fundamentally different from that of systems with two degrees of freedom. In contrast to lower dimensional cases, the nonlinear resonances form a connected network known as the Arnold web, with resonance junctions formed by the intersection of two or more independent resonances. Such a feature cannot manifest in systems with less than three degrees of freedom\cite{wigbook}. The globally connected stochastic layers along with the various junctions can result in a  form of long time phase space transport known as Arnold diffusion. Although the timescales for Arnold diffusion are possibly way too large for any observable influence, the presence of resonance junctions can lead to stability regimes called as the Nekhoroshev stability. We refer the reader to some of the earlier works for a detailed introduction to the field\cite{Cincottaetal2014,EFTHYMIOPOULOS201319}. Apart from the consequences for classical phase space transport, one other crucial aspect is to be noted in the context of trapped BECs.  Since the Arnold web is a network of nonlinear resonances, one expects on rather general grounds that the phenomenon of resonance-assisted\cite{brodieretalprl01} and chaos-assisted\cite{tomsovic94} tunneling (RAT and CAT) will also be sensitive to the features on the web. In particular, due to the substantial degeneracy of the quantum states near the junctions one can have enhanced quantum transport due to the availability of multiple pathways\cite{KarmakarJPCA2018,Firmbach_PRE_2019,PittmanJCP2016}. Therefore, combined with the fact that  chaotic dynamics occurs in the vicinity of the junctions, there is an intriguing possibility of subtle competition between the classical and quantum transport. Note that recent studies\cite{ManikandanKS,KarmakarJPCA2018,SKPKYKS2020,Karmakar_pccp_perspective} in the molecular context have started to make progress on unraveling the influence of the junctions on both the classical and quantum transport. Thus, the multiple well trapped BECs provide a possible paradigm to experimentally identify these interesting higher dimensional transport and novel stability regimes. 

\begin{figure}[tbp]
	\centering
	\includegraphics[width=0.95\linewidth]{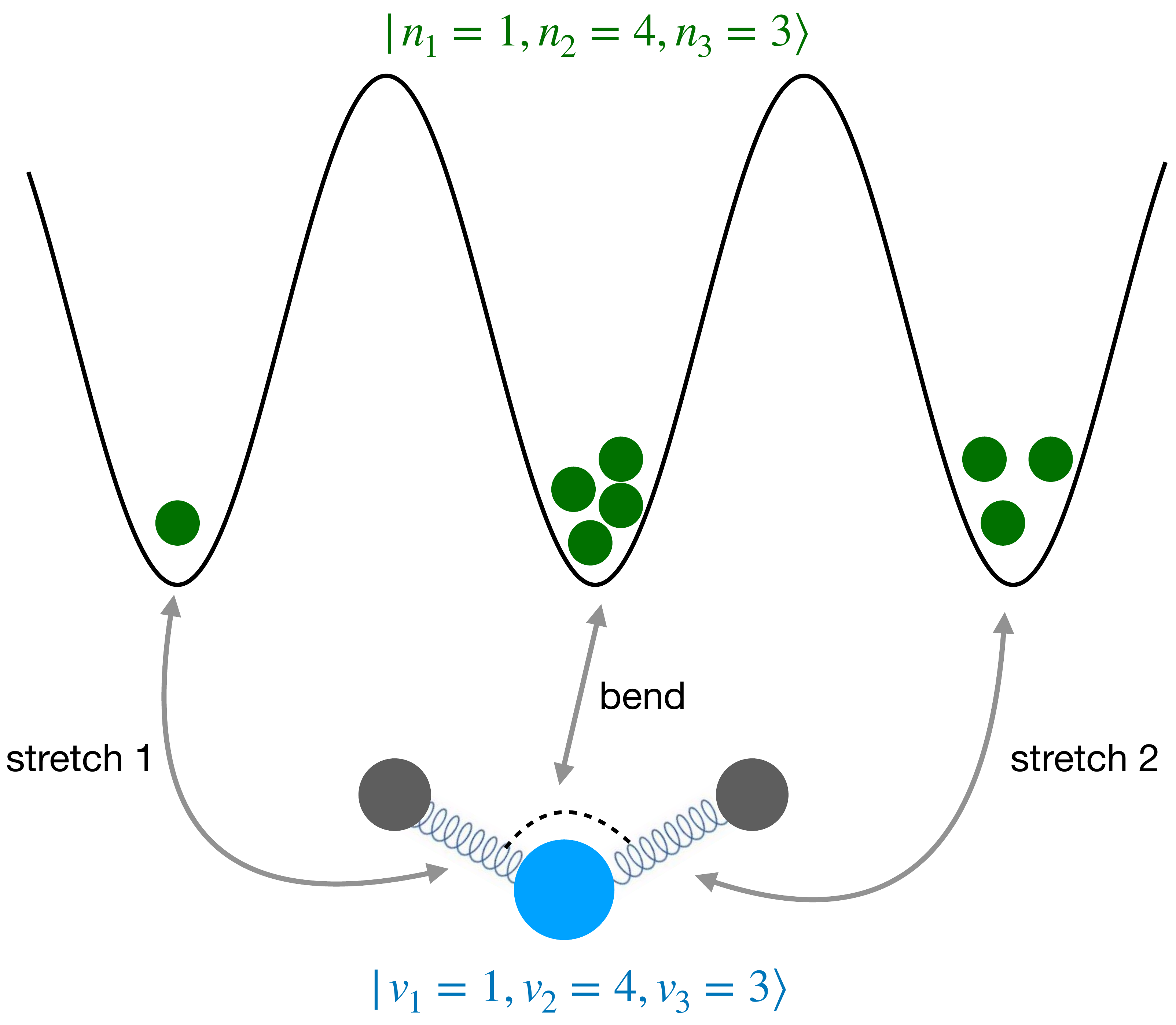} 
	\caption{A schematic illustrating the correspondence between the wells (sites) of a trapped BEC and the vibrations of a triatomic molecule. The site occupancy $n_{i}$ map to the vibrational excitations with quantum numbers $v_{i}$ and the Bose-Hubbard Hamiltonian model for BEC corresponds to the effective spectroscopic Hamiltonian for the molecule. See text for details.}
	\label{fig:bhh_vib_schematic}
\end{figure}

It is also interesting to mention a analogy between the trapped BEC systems and the vibrational dynamics of molecular systems. For several decades, 
molecular spectroscopists have been dealing
with effective Hamiltonians in order to correctly
describe the spectra and dynamics of highly excited
vibrational states of molecules. The effective Hamiltonian in this
instance is also an example of a BHH, only with the sites being
the vibrational modes and the particles (site occupancies) being the number of
vibrational excitation quanta (see Fig.~\ref{fig:bhh_vib_schematic}). Nonlinearities in the molecular
context arise due to the intrinsic anharmonicity of the bonds
whose correct description is paramount to assigning the spectra.
The nonlinearities play a crucial
role in the dynamics and are also responsible, via
different types of bifurcations, for several novel
modes of vibration that can appear or disappear in a 
vibrationally excited molecule\cite{Kellman2007,Farantos2009}. For instance, the classical bifurcation resulting in the Rabi to Josephson transition\cite{ZiboldPRL2010} in trapped BEC corresponds to a normal to local mode transition in molecules.
In particular, a large body of theoretical work has focused on
understanding and assigning the complicated quantum eigenstates
from a classical-quantum correspondence viewpoint\cite{manikandan2009}. Such analysis
are very important for understanding the phenomenon of intramolecular
vibrational energy redistribution in molecules and ultimately
lead to the formulation of accurate theories of reaction dynamics\cite{MGPGWACR2004,srihariIVR2013,Karmakar_pccp_perspective,Leitner2015}.

In this article we study a particular four well trapped BEC system that has been analyzed in detail by Khripkov, Cohen, and Vardi\cite{KhripkovJPCA2016,KhripkovNJP2015,KhripkovPRE2018}.  The system can be thought of as a linear trimer BEC coupled weakly to a fourth well or monomer. We investigate the influence of the weakly coupled monomer on the trimer subsystem by mapping out the Arnold web of the system. This allows us to highlight the competition between the classical and quantum transport in terms of the dynamics and stability of certain initial trimer configurations. In particular, we show that the quantum dynamics senses the classical Arnold web and illustrate the influence of a resonance junction on the fragmentation dynamics of the BEC. 

\section{Theory and Methods}

\subsection{The Model Hamiltonian: Quantum aspects}
\label{subsect:model}

\begin{figure*}
    \centering
    \includegraphics[width=1.0\linewidth]{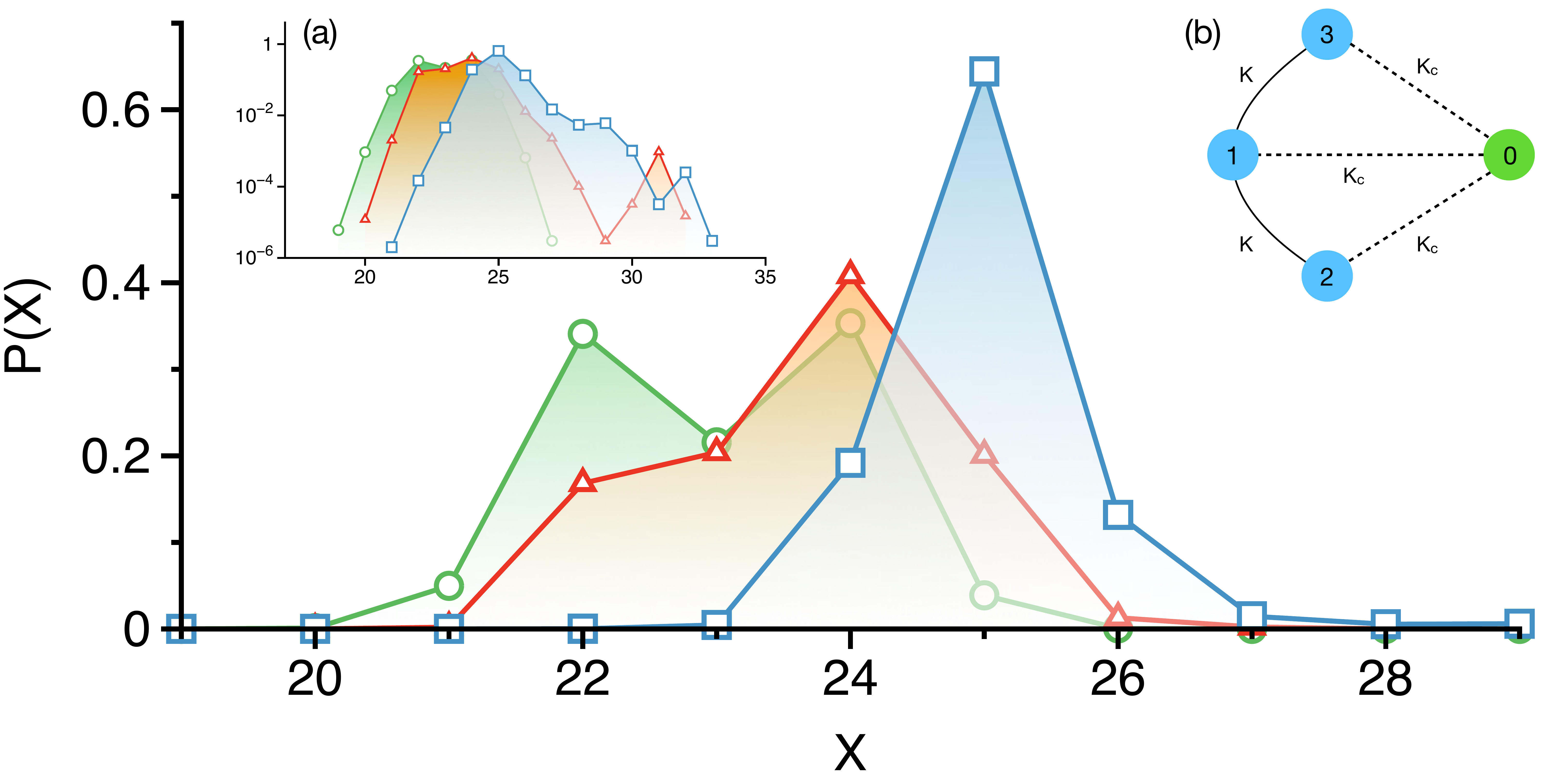}
    \caption{The distribution of $x = n_{1} + n_{2} + n_{3}$ at time $Kt = 1000$ for three example trimer eigenstates  upon coupling the monomer. The selected trimer eigenstate belong to the $x = 25$ (blue squares), $x = 24$ (red triangles) and $x = 23$ (green circles) manifold. (a) Inset showing the same plot on a log-scale. (b) A schematic of the four-site Bose-Hubbard model with the site numbering used in the text. Value of the parameters used are $U=0.5, K=0.1, K_{c}=0.05$, and $N=40$.}
    \label{fig:fig1}
\end{figure*}

We consider a four site BHM consisting of a linear trimer chain weakly coupled to a monomer. A schematic of the model is shown in Fig.~\ref{fig:fig1}(b) and such minimal  models have been studied in detail by Khripkov, Cohen and Vardi\cite{KhripkovJPCA2016} to understand the mechanism of onset of thermalization in bipartite systems. We utilize the same Hamiltonian as in these earlier studies and the model Hamiltonian can be expressed as 
\begin{eqnarray}\label{H3wqm}
 \hat{H} &=& \frac{U}{2}\sum^3_{j=1}\hat{n}_j^2 - \frac{K}{2}\left(\hat{a}_1^{\dagger}\hat{a}_2 + \hat{a}_1^{\dagger}\hat{a}_3 + {\rm h.c.}\right) + \hat{H}_{M} \nonumber \\
 & \equiv & \hat{H}_{T} + \hat{H}_{M}
\end{eqnarray}
The trimer Hamiltonian $\hat{H}_{T}$ pertains to the linear three-well system. The perturbing monomer interaction Hamiltonian is given by
\begin{equation}\label{Hpqm}
 \hat{H}_{M} =  \frac{U}{2} \hat{n}_{0}^{2} - \frac{K_c}{2}\sum^3_{j=1} \left(\hat{a}_j^{\dagger}\hat{a}_0 + {\rm h.c.}\right). 
\end{equation}
The bosonic operators, $\hat{a}_j$, $\hat{a}_j^{\dagger}$ and $\hat{n}_j = \hat{a}_j^{\dagger} \hat{a}_j$ are the annihilation, creation and number operator respectively associated with the $j^{\rm th}$ well. The site energies (interaction between particles in a well) are denoted by $U$  and $K$ is the interwell hopping (also called the tunneling) term that couples the trimer wells. The strength $K_{c}$ in  $\hat{H}_{M}$ is the  coupling strength between the trimer system and the monomer. In particular a finite value of $K_{c}$ can cause particle exchange between the trimer and the monomer wells. 

It is useful to define the trimer population  $x = \sum_{i=1}^3 n_i$. In the absence of the monomer-trimer coupling i.e., $K_{c}=0$ one has $[\hat{H},\hat{x}] = 0$. Therefore, the trimer population is a conserved quantity. However, with $K_c \ne 0$  the trimer population $x$ is no longer conserved. Nevertheless,  the total particle number $N = x + n_0 = \sum^3_{j=0} n_j$ commutes with the Hamiltonian  $[\hat{H},\hat{N}] = 0$, and is therefore a global  conserved quantity for the coupled monomer-trimer system. At this stage note that the dynamical regime of the system is captured by the following dimensionless parameters
\begin{equation}
    u = \frac{xU}{K} \,\,\,\,\, ; \,\,\,\,\, k = \frac{K_{c}}{K}
\end{equation}
Note that the effective Planck constant for the system is then $N^{-1}$ and the trimer subsystem can be classified as being in the Rabi ($ u < 1$), Josephson ($1 < u < N^{2}$), and Fock ($ u > N^{2}$) regimes. 
The parameters used in this work are $N = 40$, $U = 0.5$, $K = 0.1$, and $K_c = 0.05$. This choice of parameters implies the scaled parameter values of $u = 5x$ and $k=0.5$ and hence the system is expected to be in the Josephson regime. Apart from the fact that similar parameters have been used in  earlier studies, as seen in the next section, the strong on-site interaction results in a weakly perturbed classical Hamiltonian. The classical phase space of such  Hamiltonians generically  have mixed regular and chaotic features and therefore a rather structured Arnold web.

The quantum dynamics of an initial nonstationary state can be calculated using the eigenstates and eigenvalues of the full quantum Hamiltonian. Specifically, due to the constancy of $N$ (the total particle number) the Hamiltonian matrix is block-diagonal with a finite Hilbert space dimension of $(N+1)(N+2)(N+3)/6$.  The eigenstates $|\Psi_{\alpha}\rangle$ with eigenenergies $E_{\alpha}$ are expressed in terms of the number (Fock) basis as
\begin{equation}
    |\Psi_{\alpha}\rangle = \sum_{{\bf n} \in N} C_{{\bf n}\alpha} |n_{1},n_{2},n_{3};N\rangle
\end{equation}
In the above we use the notation $|{\bf n}\rangle \equiv |n_{1},n_{2},n_{3};N\rangle \equiv |n_{0}=N-n_{1}-n_{2}-n_{3},n_{1},n_{2},n_{3}\rangle$ for simplicity. Note that in our computations we use a symmetric basis as in earlier studies\cite{KhripkovNJP2015} on the same model system. 
A measure for the extent of localization of an eigenstate in the Fock basis is the inverse participation ratio (IPR)
\begin{equation}
    L_{\alpha} \equiv \sum_{{\bf n} \in N} \left|C_{{\bf n}\alpha}\right|^{4}
\end{equation}
For states that are completely localized or delocalized the limiting values of IPR are $\approx 1$ and $\approx N^{-1}$ respectively. Conversely, an initial Fock state's time-dependence is computed via
\begin{equation}
    |{\bf n}(t)\rangle = \sum_{\alpha} e^{-iE_{\alpha}t} C_{{\bf n}\alpha} |\Psi_{\alpha}\rangle
\end{equation}
and a measure for the number of eigenstates that contribute to the dynamics of the initial state is gauged by the dilution factor defined as
\begin{equation}
    \sigma_{\bf n} \equiv \sum_{\alpha} \left|C_{{\bf n}\alpha}\right|^{4}
\end{equation}
Again, low (high) values of $\sigma_{\bf n}$ signal a large (small) number of eigenstates that participate in the dynamics of the initial state of interest. 

An interesting question to ask at this stage, from a thermalization perspective, is the fate of a trimer eigenstate upon coupling of the monomer. Note that the trimer eigenstate, characterized by a constant $x_{0}$, in the absence of the monomer-trimer coupling ($K_{c}=0$) is a stationary state and the dynamics remains confined to the constant $x = x_{0}$ manifold. However, for $K_{c} \neq 0$ $x$ is no longer conserved. As a consequence the trimer eigenstate that initially belongs to a specific $x$-manifold will mix with the trimer eigenstates belonging to the different $x$-manifolds. In other words, the uncoupled delta distributed $\delta(x-x_{0})$ will now spread and is described by a distribution $P(x,x_{0};t)$ that evolves with time. In Figure~\ref{fig:fig1} we show such a mixing among the different $x$-manifolds  for one trimer eigenstate each belonging initially to the $x = 23, 24$, and $25$ manifold. The observed long-time spreading along $x$ gauges the extent to which the monomer-trimer coupling perturbs the trimer dynamics.  In Fig.~\ref{fig:fig1}(a) the mixing of the different $x$-manifolds is shown on the log-scale. Clearly, fig.~\ref{fig:fig1}(b) reveals subtle aspects of the mixing that are not apparent from inspecting the distribution on a linear scale.  Note that, not all the trimer eigenstates show influence of the coupling and there do exist trimer eigenstates that essentially  remain unperturbed. Following Khripkov et al\cite{KhripkovJPCA2016} it is possible to understand the distribution $P(x,x_{0};t)$ by studying the associated Fokker-Planck equation and comparing to the classical ergodic predictions. However, we do not take this approach and instead show that the observed differences in spreading in Fig.~\ref{fig:fig1} is closely related to the features in the classical Arnold web and their implication for the quantum dynamics.

\subsubsection{Measures for gauging the influence of the monomer on the trimer dynamics}

In addition, for the specific model of interest, it is useful to further define two quantities. The first one has to do with a measure of the influence of the monomer perturbation in eq.~\ref{Hpqm} on the dynamics of the initial states belonging to the trimer subsystem. Generally, the dilution factor of a Fock state for the uncoupled system ($K_c = 0$) and the coupled system ($K_c \ne 0$) are expected to be different. To extract the influence of the monomer, we define dilution factor ratio (DFR) $\widetilde{\sigma}_{\mathbf{n}}$ as
\begin{equation}\label{eq:dfr}
\widetilde{\sigma}_{\mathbf{n}} = \frac{\sigma_{\mathbf{n}}(K_c \ne 0)}{\sigma_{\mathbf{n}}(K_c = 0)} 
\end{equation}
By definition, the trimer Fock states which are influenced by the presence of the monomer will have $\widetilde{\sigma}_{\mathbf{n}} < 1$ whereas for the protected trimer states $\widetilde{\sigma}_{\mathbf{n}} \approx 1$  is expected. The former case leads to a strong breakdown of the constancy of $x$ and implies particles exchange between the trimer and the monomer. A second set of measures, which are  experimentally measurable in trapped BEC systems, are the various population imbalances. For the trimer subsystem, they\cite{MunroPRA2000} can be defined as 
\begin{eqnarray} \label{eq:imbeqn}
X_1 & = & \frac{(n_1-n_3)}{N} \\
X_2 & = &\frac{1}{3} \frac{(n_1 + n_3 - 2n_2)}{N} \nonumber
\end{eqnarray}
Computing $X_{1}$ and $X_{2}$ as a function of time for  $K_{c} \neq 0$ and comparing the results to the uncoupled trimer case $K_{c}=0$ is expected to reveal the extent to which the monomer influences the dynamics of the initial Fock state of interest.
In what follows, the DFR and population imbalance measures will be used to understand the mechanism of $x$-mixing and the location of protected trimer states.

\subsection{Classical limit Hamiltonian: Nonlinear resonances and the zeroth-order Arnold web}
\label{subsect:cmham}

The classical analogue of the quantum Hamiltonian in eq. \ref{H3wqm} is obtained by using the Heisenberg correspondence
\begin{equation}
 \hat{a}_{j} \longleftrightarrow \sqrt I_j \hspace{1mm} e^{i\theta_j}\,\,,\,\, \hat{a}^{\dagger}_{j} \longleftrightarrow \sqrt I_j \hspace{1mm} e^{-i\theta_j}
\end{equation}
for the j$^{\rm th}$-site where, $({\bf I}, {\bm \theta})$ are the classical action and angle variable respectively. The corresponding classical Hamiltonian  can be expressed in action-angle variables as
\begin{equation} \label{H3wcl}
{\cal H}({\bf I},{\bm \theta})  
\equiv {\cal H}_{T}({\bf I},{\bm \theta}) + {\cal H}_{M}({\bf I},{\bm \theta})
\end{equation}
with the trimer Hamiltonian being
\begin{equation}
    {\cal H}_{T}({\bf I},{\bm \theta}) = \frac{U}{2}\sum^3_{j=1}I_j^2 - K \sum_{j=2,3} \sqrt{I_{1}I_{j}} \cos \theta_{1j}
\end{equation}
and the perturbing monomer Hamiltonian corresponding to eq.~\ref{Hpqm} 
given by
\begin{equation}\label{Hpcl}
 {\cal H}_{M}({\bf I},{\bm \theta}) = \frac{U}{2}I_0^2 - K_c \sum^{3}_{j=1}\sqrt{I_{0} I_{j}} \cos \theta_{0j}
\end{equation}
In the above we have used the notation $\theta_{kl} \equiv \theta_{k} - \theta_{l}$. Note that even with $K_{c}=0$ the trimer Hamiltonian ${\cal H}_{T}$ is classically non-integrable. The Hamiltonian ${\cal H}({\bf I},{\bm \theta})$ has the form of a typical multidimensional resonant Hamiltonian and, in analogy to the quantum case, one can explicitly show that the quantity
\begin{equation}
    {\cal N} = \sum_{j=0}^{3} I_{j} \equiv x_{c} + I_{0}
\end{equation}
is a conserved quantity i.e., the Poisson bracket $\{{\cal H},{\cal N}\}=0$. In other words, ${\cal N}$ is a constant of the motion. Similarly, analogous to the quantum  $\hat{x}$, the quantity $x_{c} = \sum_{j=1}^{3} I_{j}$ is conserved for $K_{c}=0$.

\begin{figure*}[htbp]
\centering
	\includegraphics[width=0.95\linewidth]{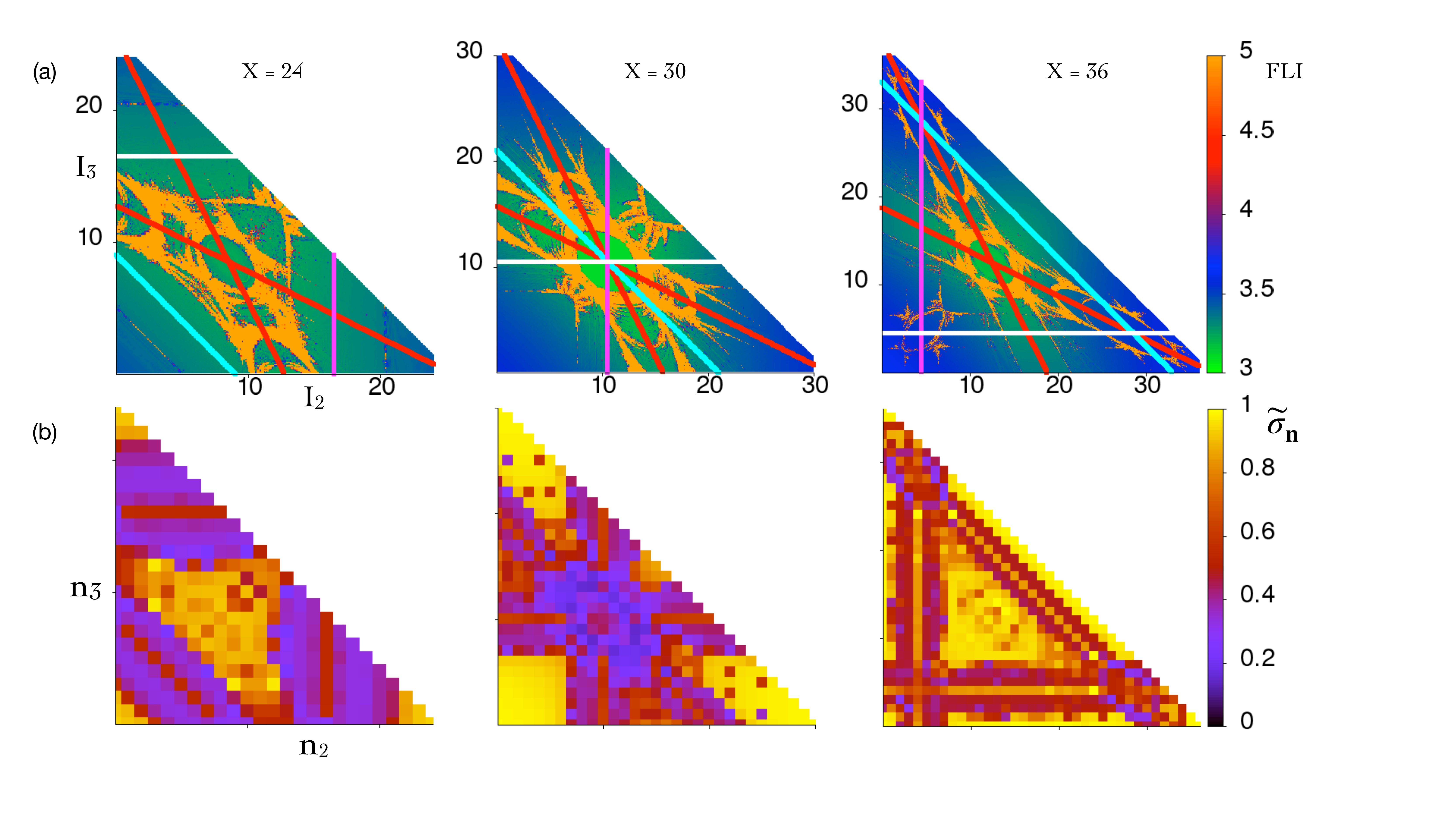} 
	\caption{ \label{fig:df} (a) The classical Arnold web projected onto the $(I_{2},I_{3})$ space for three different values of the trimer population $x$ with total number of particles $N=40$. Note that FLI values $\geq 5$ are color coded as yellow. (b) The corresponding quantum dilution factor ratio $\tilde{\sigma}_{\mathbf{n}}$ of eq.~\ref{eq:dfr} shown projected onto the $(n_{2},n_{3})$ space. In (a) the zeroth-order prediction of the various resonances are also shown. The two trimer resonances $R_{Tj}$   are indicated as red lines. The three monomer-trimer resonances $R_{Mj}$ for $j=1,2,3$ are shown as cyan, magenta, and white lines respectively.}
\end{figure*}

Since ${\cal N}$ is a globally conserved quantity, the full coupled trimer-monomer Hamiltonian can be reduced via a suitable canonical transformation to an effectively three degrees of freedom system. Moreover, the fixed points of the reduced Hamiltonian will correspond to periodic orbits of the original system and it is possible to analyze the stability of the periodic orbits and their bifurcations with varying system parameters. However, we do not undertake  this task in the present work and instead focus on the structure of the Arnold web and its influence on the classical and quantum dynamics. For further analysis it is useful to rewrite  the Hamiltonian ${\cal H}({\bf I},{\bm \theta})$ of eq.~\ref{H3wcl} in the form of a typical multidimensional resonant Hamiltonian
\begin{equation} \label{multiresham}
    {\cal H}({\bf I},{\bm \theta}) = {\cal H}_{0}({\bf I}) + V({\bf I},{\bm \theta})
\end{equation}
with ${\cal H}_{0}({\bf I}) \equiv U \sum_{j=0}^{3} I_{j}^{2}/2$ and
\begin{equation}
    V({\bf I},{\bm \theta}) \equiv - K \sum_{j=2,3} \sqrt{I_{1}I_{j}} \cos \theta_{1j} - K_c \sum^{3}_{j=1}\sqrt{I_{0} I_{j}} \cos \theta_{0j}
\end{equation}
One can now define the zeroth-order frequencies
\begin{equation}
    {\bm \Omega}({\bf I}) = \frac{\partial {\cal H}_{0}({\bf I})}{\partial {\bf I}}
\end{equation}
and the location of the various resonances and their Chirikov widths in the ${\bf I}$ space can be determined. From the form of the  perturbation term $V({\bf I},{\bm \theta})$ above it is clear that the various actions are all involved in five pairwise $1:1$ nonlinear resonances. The resonant surfaces can then be projected onto a convenient two-dimensional action subspace. For example, the five resonances can be expressed in $(I_{2},I_{3})$ space as 
\begin{eqnarray}
\Omega_{0}({\bf I}) &=& \Omega_{1}({\bf I}) \implies R_{M1}: I_{3} = (2x_{c}-{\cal N}) - I_{2} \nonumber \\
\Omega_{0}({\bf I}) &=& \Omega_{2}({\bf I}) \implies R_{M2}: I_{2} = {\cal N} - x_{c} \nonumber \\
\Omega_{0}({\bf I}) &=& \Omega_{3}({\bf I}) \implies R_{M3}: I_{3} = {\cal N} - x_{c} \nonumber \\
\Omega_{1}({\bf I}) &=& \Omega_{2}({\bf I}) \implies R_{T1}: I_{3} = x_{c} - 2 I_{2} \nonumber \\
\Omega_{1}({\bf I}) &=& \Omega_{3}({\bf I}) \implies R_{T2}: 2I_{3} = x_{c} - I_{2} \nonumber
\end{eqnarray}
In the above $R_{Mj}$ and $R_{Tj}$ stand for the three monomer-trimer and the two trimer only resonances with widths scaling as $\sqrt{K_{c}}$ and $\sqrt{K}$ respectively. The disposition of these resonances along with their intersections for a given $(x_{c},{\cal N})$ forms the Arnold web. 

Note that for a fixed $(x_{c},{\cal N})$ it is not necessary that all the resonances intersect to generate  multiplicity-$2$ resonance junctions in the action subspace of choice. In fact, it can be shown that the $R_{Mj}$ resonances will intersect in the $(I_{2},I_{3})$ space for $x_{c} \geq 2{\cal N}/3$. In other words only if the trimer population is sufficiently large then resonance junctions manifest in the$(I_2,I_3)$ space. However, intersections can happen in a different projection and an example of this can be seen later. Moreover, due to the constraint $I_{2} + I_{3} \leq x_{c}$, in the $(I_{2},I_{3})$ space the projected Arnold web is restricted to a triangular region with $I_{2}^{\rm max} = I_{3}^{\rm max} = x_{c}$. The constraint arises due to the fact that semiclassically the actions correspond to quantum populations which are necessarily positive. It is also possible to locate the various multiplicity-$2$ junctions on the Arnold web from the zeroth-order arguments. In particular it can be shown that all five resonances intersect at $(I_{2}^{\times},I_{3}^{\times}) = ({\cal N}/4,{\cal N}/4)$ provided $x_{c} = 3 {\cal N}/4$.

\subsubsection{Computation of the Arnold web}

The  description of the Arnold web in the previous section is expected to be reasonably accurate for weakly coupled systems. However, the zeroth-order arguments cannot yield the correct resonance widths and the nature of the dynamics near different resonance junctions. The zeroth-order web, representing the ``geography" of the resonance network up to some finite order, does not necessarily indicate the dynamical dominance of specific resonances either\cite{SKPKYKS2020}. Consequently, it is important to map out the true Arnold web for the system at the parameters of interest. Therefore, we now explicitly map the classical Arnold web  for the system of interest using the technique of fast Lyapunov indicator (FLI)\cite{Froeschle1997,FroeschleScience2000}.The details of the FLI technique and its implementation have been discussed in several earlier works\cite{chaosdetect}. Therefore, we provide a brief description below and refer to the published literature for 
further details. Note that one can use several other techniques\cite{SKOKOS200730,Skokos_2004,BARRIO2005711,CINCOTTA2003151} as well to compute the Arnold web including the recently proposed approach based on Shannon entropy\cite{CINCOTTA2021132816}. 

For a general dynamical system 
\begin{equation}
    \frac{d\mathbf{X}}{dt} = \mathbf{F}(\mathbf{X})
    \label{flieom}
\end{equation}
with $\mathbf{X} = \{x_1, x_2, \ldots, x_i, \ldots\}$ being the dynamical variables, we compute the FLI using the definition 
\begin{equation}
    \mathrm{FLI}(\mathbf{X}(0), \mathbf{v}(0), t) = \sup_{0 < t^{'} <t} \log \Vert\mathbf{v}(t^{'}) \Vert
    \label{flitgtvec}
\end{equation}
where, $\mathbf{v}$ is a tangent vector satisfying the variational equation 
\begin{equation}
    \frac{d\mathbf{v}}{dt} = \left(\frac{\partial \mathbf{F}}{\partial \mathbf{X}} \right) \mathbf{v}
\end{equation}
The equations of motion eq.~\ref{flieom} and the variational equation eq.~\ref{flitgtvec} are integrated together to determine the FLI value associated with a specific initial condition. In addition, instead of using a fixed initial tangent vector, following Barrio et al\cite{BARRIO20091697} we choose the initial tangent vector as $-\nabla H/||\nabla H||$.

To construct the Arnold web for a given $x_{c}$, the initial angles in eq.~\ref{multiresham} are fixed to a particular value, $(\theta_1^{(0)}, \theta_2^{(0)}, \theta_3^{(0)}, \theta_0^{(0)}) = (0,0,0,0)$ and a $500 \times 500$ grid of initial actions $(I_2^{(0)}, I_3^{(0)})$ is constructed. The other actions are obtained from the conserved quantity, $I_0^{(0)} = {\cal N} - x_{c}$ and $I_1^{(0)} = x_{c} - I_2^{(0)} - I_3^{(0)}$. The initial conditions are then propagated for a total time $t_f = 5000$ units (corresponding to $Kt_f = 500$). The FLI values at the final time are projected on to the $(I_2,I_3)$ space. Note that equivalent representations can be obtained with  different choices of the initial angles and the action subspace of interest. Furthermore, the final time considered here are fairly long and larger values of $t_{f}$ do not significantly alter the features of the computed Arnold web.

\section{Results and discussions}
\label{sect:results}

\subsection{Arnold web: classical-quantum correspondence}

In Fig.~\ref{fig:df}(a), we show the Arnold web for three different values of the trimer population $x$ (equivalently $x_{c}$). These values are selected to show the web structure for $x_{c} < 2{\cal N}/3$ (no intersection of the $R_{Mj}$), $x_{c} = 1.125 (2{\cal N}/3) \equiv 3 {\cal N}/4$ (all resonances intersect), and $x_{c} > 2 {\cal N}/3$ (several distinct resonance junctions). In Fig.~\ref{fig:df}(a) the center of the resonances predicted from the zeroth order Hamiltonian are also shown superimposed on the FLI-based web. Clearly the five nonlinear resonances are  visible and the agreement between the zeroth-order prediction and the computed web are fairly good. Note that the FLI values are color coded such that the orange regions correspond to chaotic regions while the blue and green regions indicate the regular regions. As expected chaotic regions are observed  in the vicinity of the resonance junctions. Specifically, at junctions formed by the intersection of $R_{Tj}$ and $R_{Mj}$ resonances one expects a significant perturbation of the trimer population due to the monomer. It is also worth observing from Fig.~\ref{fig:df}(a) that with increasing $x_{c}$ the $R_{Mj}$ resonances sort of sweep through the $(I_{2},I_{3})$ space, intersecting the $R_{Tj}$ at $({\cal N}/4,{\cal N}/4) = (10,10)$, and finally disappearing for sufficiently large $x_{c}$.

Given the way in which the Arnold webs in Fig.~\ref{fig:df}(a) are constructed, it is natural to enquire if the features of the web influence the dynamics of the initial Fock states $|n_{1},n_{2},n_{3};N\rangle$. Towards this end in Fig.~\ref{fig:df}(b) we show the DFR, defined in eq.~\ref{eq:dfr}, associated with the Fock states. In order to compare with the Arnold web the DFRs are projected onto the corresponding $(n_{2},n_{3})$ quantum occupation number space. The classical-quantum correspondence between the classical Arnold web features and the quantum DFR values is striking. In particular, one can clearly see the DFRs reflecting the geography of the resonance network. Such a detailed correspondence has been seen earlier in a different context\cite{KarmakarJPCA2018}. It is evident from Fig.~\ref{fig:df}(b) that the location of the fragmented Fock states ($\widetilde{\sigma}_\mathbf{n} < 1$) due to the monomer-trimer perturbation are exactly where the resonances involving the monomer are present. For instance, Fig.~\ref{fig:df}(a) shows that for $x = 24, 36$, only the trimer resonances are present in the region around $I_1 \approx I_2 \approx I_3$ and the resonances involving the monomer are away from that region. However, for $x = 30$, all the resonances intersect at $I_1 = I_2 = I_3 = x/3 \equiv {\cal N}/4$. Consequently, for $x = 24$ and $36$  around $I_1 \approx I_2 \approx I_3$ the presence of only trimer resonances  causes transport (fragmentation) within the trimer subsystem i.e., $\widetilde{\sigma}_\mathbf{n} \approx 1$. In other words such trimer configurations are protected from the monomer-trimer perturbation. On the other hand, for $x=30$ the intersection of all the resonances around $I_1 \approx I_2 \approx I_3$ leads to particle exchange among all the four wells and therefore $\tilde{\sigma}_\mathbf{n} < 1$. In this case the trimer configuration $|N/3,N/3,N/3;N\rangle$ is no longer protected. It is also interesting to note that trimer configurations of the form $|0,x,0;N\rangle$ and $|0,0,x;N\rangle$, components of the so called ``NOON states", are protected for $x=30$ but not so much for $x=24$ and $x=36$. 

We end this section by making a few observations. Firstly, along a single resonance in Fig.~\ref{fig:df}(a), and away from a junction, the particle exchange within the two wells involved is similar to the typical two well trapped BEC dynamics. Thus, one may anticipate the existence of MQST states. However, near a resonance junction the presence of multiple resonances involves three or more wells in the dynamics. The particular wells which are actively involved in the fragmentation of the BEC are  identifiable from the Arnold web. 
The excellent correspondence between the DFRs and the Arnold web features therefore suggests that the stability of specific trimer configurations of interest  can be predicted by referring to the underlying classical Arnold web.  Secondly,  a closer inspection of Fig.~\ref{fig:df}(b) reveals subtle variations in the DFR even in the (un)protected regions. The origins for such behaviour has to do with the phenomenon of CAT\cite{bohigas93,tomsovic94,eltschka05} and RAT\cite{brodieretalprl01,eltschka05,ksjcp05} in the vicinity of the multiplicity-$2$ resonance junctions and has been brought out clearly in  earlier works\cite{KarmakarJPCA2018,Karmakar_pccp_perspective,KSPRE2005} on  different model systems. Near a resonance junction one expects a competition between transport due to local classical chaos and quantum transport dressed due to dynamical tunneling among near-degenerate states. In fact, hints of a Nekhoroshev type stability can be seen in Fig.~\ref{fig:df}(b) even for $x=30$. We anticipate that these observations are connected to the work\cite{KhripkovNJP2015} of Khripkov et al wherein they argue that the glassiness of the quantum network of transitions leads to a percolation-like dynamics and slow thermalization in model bipartite systems. Similar arguments and observations have been made nearly two decades ago in the context of IVR in molecules and the persistence of nonstatistical reaction dynamics even at fairly high excitation energies and density of states. It is also interesting to compare our Arnold webs to the recent study\cite{Nigro_PRA_2018} by Nigro et al wherein a ring shaped four-site trap potential is analyzed. 

\subsection{Fock state dynamics and the Arnold web: dynamical tunneling and resonance junctions}

A comparison of Fig.~\ref{fig:df}(a) and (b) reveals that much of the rich structure on the Arnold web does influence the quantum dynamics of the BEC. Indeed, with increasing particle number $N$ (decreasing $\hbar_{\rm eff} \sim 1/N$) even the tiniest features on the Arnold web should be sensed by the quantum dynamics. Therefore, a detailed study of various classes of initial trimer states with the Arnold web as a template is certainly worth investigating. However, as a preliminary effort, in the current study we illustrate the interplay of classical and quantum transport induced by features on the Arnold web for three example Fock states. The states of interest are $|15,0,10;40\rangle$, $|16,0,8;40\rangle$, and $|17,0,6;40\rangle$. The motivation for the specific states considered is threefold. Firstly, these states correspond to initial configurations with one of the trimer wells being empty. In the context of trapped BEC the dynamics of such  single well depleted states is of interest. A second motivation has to do with the fact that the initial trimer eigenstates utilized in Fig.~\ref{fig:fig1} have  dominant contribution from the Fock states of interest. Thus, the expectation is that the long time distribution $P(x)$ observed in Fig.~\ref{fig:fig1} can be rationalized based on the corresponding Fock state  dynamics in the presence of the monomer-trimer perturbation. Thirdly, as shown and discussed below, the chosen set of states approach a  junction on the Arnold web, resulting in fragmentation dynamics with a subtle mix of classical and quantum mechanisms.   

\begin{figure}[h]
	\centering
	\includegraphics[width=0.85\linewidth]{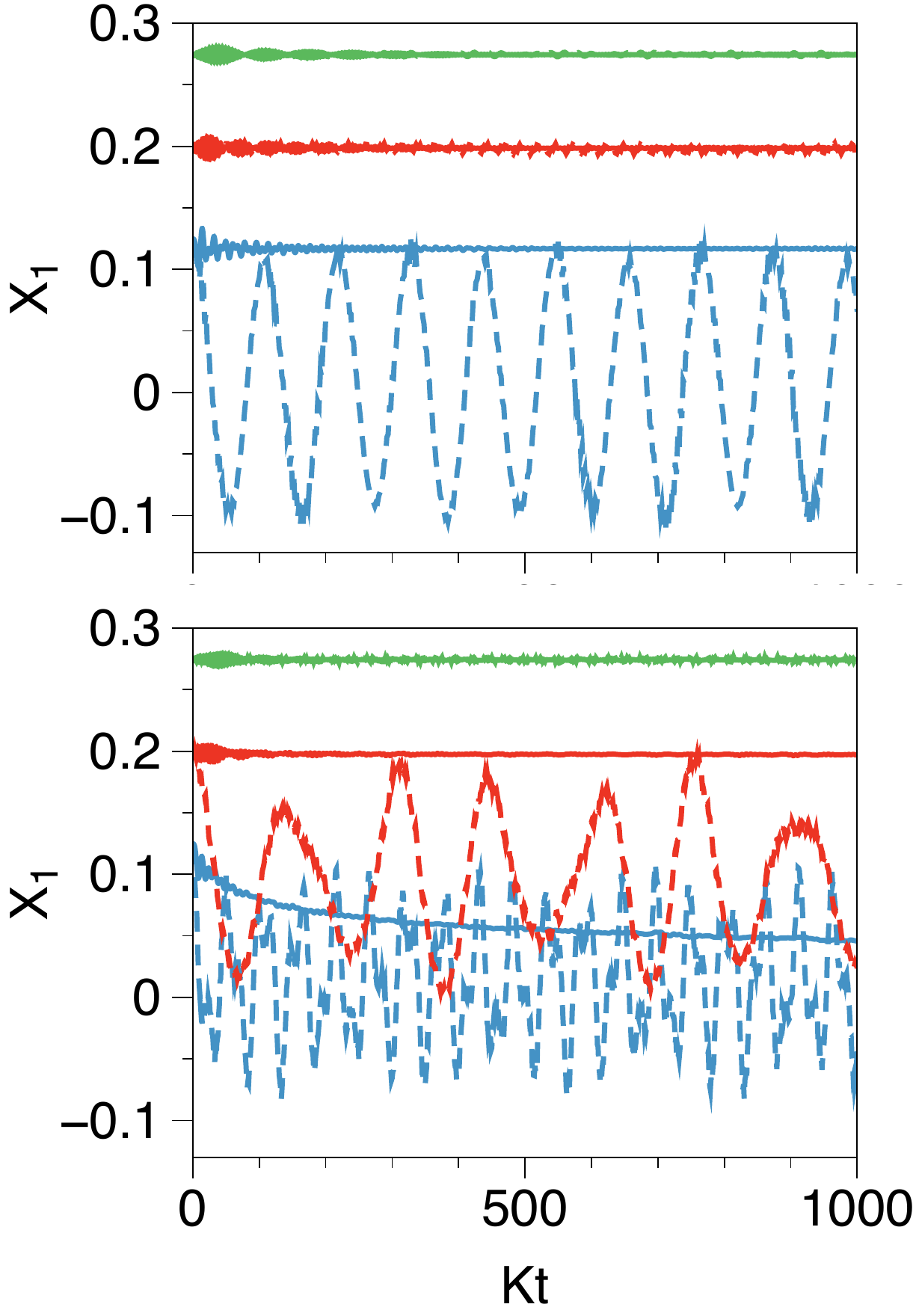} 
	\caption{ \label{fig:imb} The average quantum (dashed line) and classical (solid line) population imbalance $X_{1}$ (cf. eq.~\ref{eq:imbeqn}) as a function of time for the Fock states $|15,0,10;40\rangle$ (blue), $|16,0,8;40\rangle$ (red) and $|17,0,6;40\rangle$ (green). (Top panel) Monomer uncoupled ($K_{c}=0$) and (Bottom panel) Monomer coupled ($K_{c} \neq 0$. Note that the colors assigned to the Fock states are consistent with the colors in Fig.~\ref{fig:fig1} associated with the trimer eigenstates. Thus, for instance, the Fock state $|15,0,10;40\rangle$ contributes dominantly to the initial trimer eigenstate used in Fig.~\ref{fig:fig1} with $x=25$.}
\end{figure}

\subsubsection{Population imbalance: quantum versus classical}

In order to set the tone for the following discussions, Fig.~\ref{fig:imb} shows the time dependence of the avearge quantum, and the analogous classical, population imbalance $X_{1}$ (see eq.~\ref{eq:imbeqn}) for the three states.  The $X_{2}$ measure in eq.~\ref{eq:imbeqn} is not shown here since there is no new information to be gained for the set of states that are of interest. This has to do with the fact that the well number two remains unoccupied, as can also be expected based on the location of the states on the Arnold web. The uncoupled $K_{c}=0$ (Fig.~\ref{fig:imb} top panel) results are compared with the coupled $K_{c} \neq 0$ (Fig.~\ref{fig:imb} bottom panel) to assess the influence of the monomer-trimer coupling. We notice distinctively different behaviours even though the states are located very closely in the Fock space.
In the case of the state $|17,0,6;40\rangle$ neither the quantum nor the classical population imbalance shows any significant decay with time irrespective of the presence or absence of the monomer coupling.  Therefore, this particular trimer configuration is fairly well protected.  In case of the 
state $|16,0,8;40\rangle$, for $K_{c}=0$ both the classical and quantum $X_{1}$ do not decay indicating that the trimer configuration is robust. However, with a finite monomer-trimer coupling  the state is protected classically  but not quantum mechanically. The quantum $X_{1}$ shows  oscillations with multiple timescales, implying that the fragmentation of this state is mainly due to a quantum mechanism. Typically, based on earlier studies, we anticipate the quantum mechanism to be associated with the phenomenon of RAT. In contrast,
the state $|15,0,10;40\rangle$ exhibits a more complicated behaviour. Even for $K_{c}=0$  there is quantum fragmentation, while it is classically protected, and the $X_{1}$ shows nearly coherent oscillations (period $Kt \sim 100$). Again, quantum dynamical tunneling within the trimer subspace is expected to be the cause of the observed fragmentation. With the monomer coupled,  the quantum $X_{1}$ shows an oscillatory behaviour, but less coherent. Interestingly, with the monomer coupled the classical $X_{1}$ also decays and hence the dynamics of $|15,0,10;40\rangle$ is presumably due to a mix of classical and quantum transport.   

\subsubsection{Survival probabilities: dynamical tunneling?}
Clearly,
understanding these three distinctively different behaviour of the population imbalance and therefore the spreading along $X$ in Fig.~\ref{fig:fig1} requires further analysis. To this end we compute the probabilities
\begin{equation}
    p_{{\bf n}' {\bf n}}(t) \equiv \left| \left\langle {\bf n}'|e^{-i\hat{H}t}|{\bf n} \right \rangle \right|^{2}
\end{equation}
where we have denoted $|{\bf n} \rangle = |n_{1},n_{2},n_{3};N\rangle$, the initial state of interest. The above quantity yields information on the flow of probability through the quantum number space and for the choice $|{\bf n}'\rangle = |{\bf n}\rangle$ we have the survival probability $p_{{\bf n}{\bf n}}(t) \equiv p_{\bf n}(t)$, also known as the autocorrelation function or the fidelity of the given state. Note that the Fourier transform of $p_{{\bf n}{\bf n}}(t)$ yields the local density of states (LDOS) which is of interest to studies on thermalization in interacting many-body systems.

\begin{figure}
	\centering
	\includegraphics[width=0.95\linewidth]{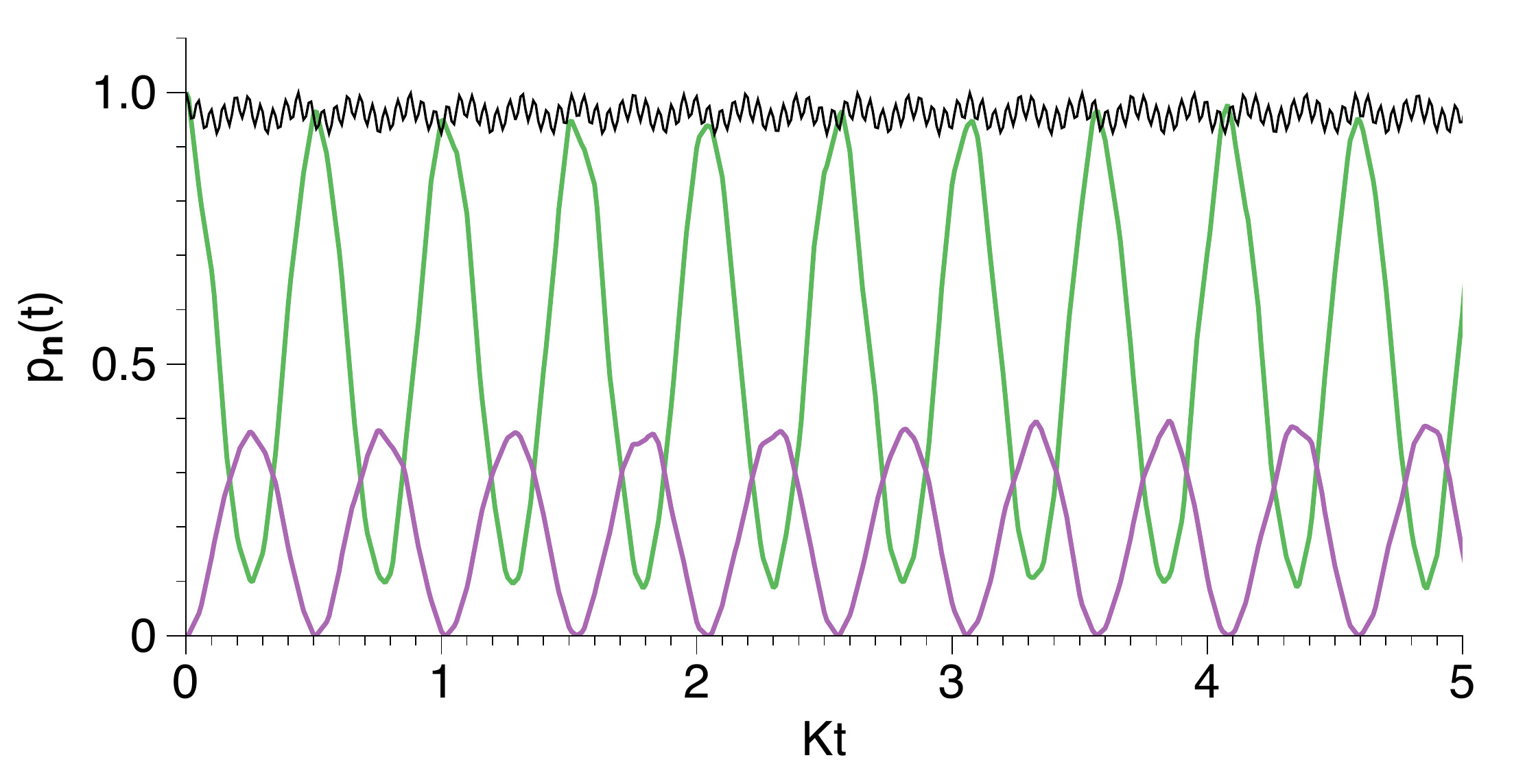} 
	\caption{Survival probability for the Fock state $|17,0,6\rangle$ for the uncoupled ($K_c = 0$, black) and coupled ($K_c =0.05$, green) cases.  Quantum  cross probability for the state $|{\bf n}'\rangle = |16,0,6;40\rangle$ (violet) for the coupled case is shown. Note that the cross probability for the state $|18,0,6;40\rangle$ (not shown) is nearly the same as that of state $|16,0,6;40\rangle$.}
	\label{fig:1706surv}
\end{figure}

First consider the fragmentation of the state $|17,0,6;40\rangle$. The relevant survival probabilities are shown in Fig.~\ref{fig:1706surv} and it is clear that in absence of the monomer the condensate does not fragment. Upon coupling the monomer the survival probability exhibits fast nearly periodic oscillations on a timescale of $Kt \sim 0.5$. The cross-survival probability analysis shows that the probability flows dominantly to the states $|n_1,0,6;40\rangle$, $n_1 = 16,18$ nearly equally. The probability flow essentially indicates the influence of the hopping term $(\hat{a}^{\dagger}_1 \hat{a}_0 + h.c.)$, which corresponds classically to the $R_{M1}$ monomer-trimer nonlinear resonance. The equivalent probability flow to states $|16,0,6;40\rangle$ and $|18,0,6;40\rangle$ indicates that on average  $\bar{n}_1 = 17$ and hence, as seen in Fig.~\ref{fig:imb}, $X_1$ does not show appreciable decay. Similar conclusions result from the classical survival probability results (not shown). Moreover, since the states correspond to $|16,0,6;40\rangle$ and $|18,0,6;40\rangle$ the trimer population $x=22$ and $x=23$ respectively, the observations support the peaks seen in Fig.~\ref{fig:fig1} for the specific trimer eigenstate distribution. As will be seen below, it is reasonable to associate the dominant features of the $x$-diffusion  in Fig.~\ref{fig:fig1} with the classical dynamics.

\begin{figure}[tbp]
	\centering
	\includegraphics[width=0.95\linewidth]{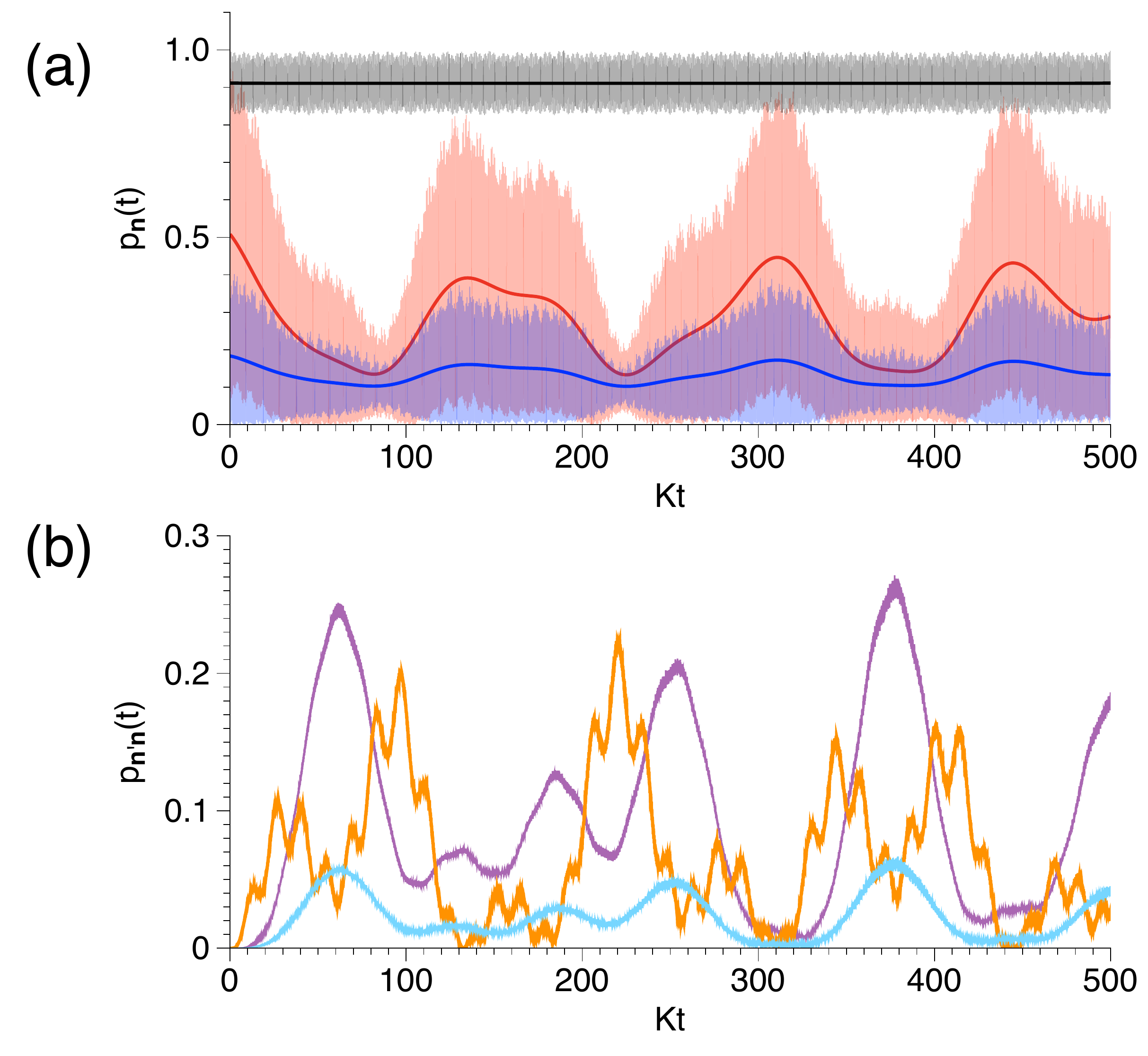} 
	\caption{(a) Survival probability for the Fock state $|16,0,8\rangle$ for the uncoupled ($K_c = 0$, black) and coupled ($K_c =0.05$, red) cases. Cross probability for the state $|15,0,8;40\rangle$ is shown in blue. Note that the smoothed probabilities, capturing the average behaviour, are shown as bold solid lines. The fast oscillations occur on a timescale $Kt \sim 0.5$.
	(b) Cross probabilities for select states involved in the dynamics of the Fock state $|16,0,8;40\rangle$ in the coupled case with $K_c =0.05$.  The states shown are $|8,0,16;40\rangle$ (violet), $|9,0,13;40\rangle$ (orange), and $|8,0,15;40\rangle$ (blue). Note that classically there is no  probability flow into these states.}
	\label{fig:1608class2prob}
\end{figure}

The fragmentation dynamics of the state $|16,0,8;40\rangle$ is 
shown in Fig.~\ref{fig:1608class2prob}(a) and compared to Fig.~\ref{fig:1706surv} the dynamics is much more complicated with the involvement of several Fock states. The short time decay (not apparent from the plot due to averaging) exhibits fast oscillations on a $Kt \sim 0.5$ timescale as before. Again, this timescale is associated with probability flow to the states $|15,0,8;40\rangle$ and $|17,0,8\rangle$ due to the $(\hat{a}^{\dagger}_1 \hat{a}_0 + h.c.)$ and hence classically associated with the location of the state in the $R_{M1}$ resonance. Thus, analogous to the discussions above, the classical $X_{1}$ in Fig.~\ref{fig:imb} with $K_{c} \neq 0$ does not vary significantly with time. On the other hand, the quantum $X_{1}$ in Fig.~\ref{fig:imb} does exhibit non-trivial time dependence and hence must be due to dynamical tunneling pathways. Indeed, as shown in Fig.~\ref{fig:1608class2prob}(b), there is considerable probability flow to a set of states that are not connected by classical pathways. The time scales seen in Fig.~\ref{fig:1608class2prob}(b) clearly correlate with the longer decay time scale ($Kt \sim 100$) seen in Fig.~\ref{fig:1608class2prob}(a) as well as in Fig.~\ref{fig:imb} for the population imbalance. It is interesting to note from Fig.~\ref{fig:1608class2prob}(b) that there is a substantial probability flow from $|16,0,8;40\rangle$ to $|8,0,16;40\rangle$ i.e., a total of eight bosons being exchanged between the trimer wells. Several potential pathways can be constructed for the same. However, as an example, the pathway $|16,0,8;40\rangle \rightarrow |15,0,8;40\rangle \rightarrow |13,0,9;40\rangle \rightarrow |9,0,13;40\rangle \rightarrow |8,0,16;40\rangle$ is particularly interesting since it involves a mix of classical and quantum mechanisms. The first and the third legs of the pathway are classically allowed. The second leg $|15,0,8;40\rangle \rightarrow |13,0,9;40\rangle$ is classically forbidden and requires a term like $\hat{a}_{1}^{2}\hat{a}_{3}^{\dagger}\hat{a}_{0}^{\dagger}$ to be feasible. Similarly, the last leg of the pathway is classically forbidden and requires a different higher order coupling term. Based on previous studies\cite{KSPRE2005,Srihari2007} it is expected that such higher order effective couplings implicate the phenomenon of RAT.

\begin{figure}[tbp]
	\centering
	\includegraphics[width=0.95\linewidth]{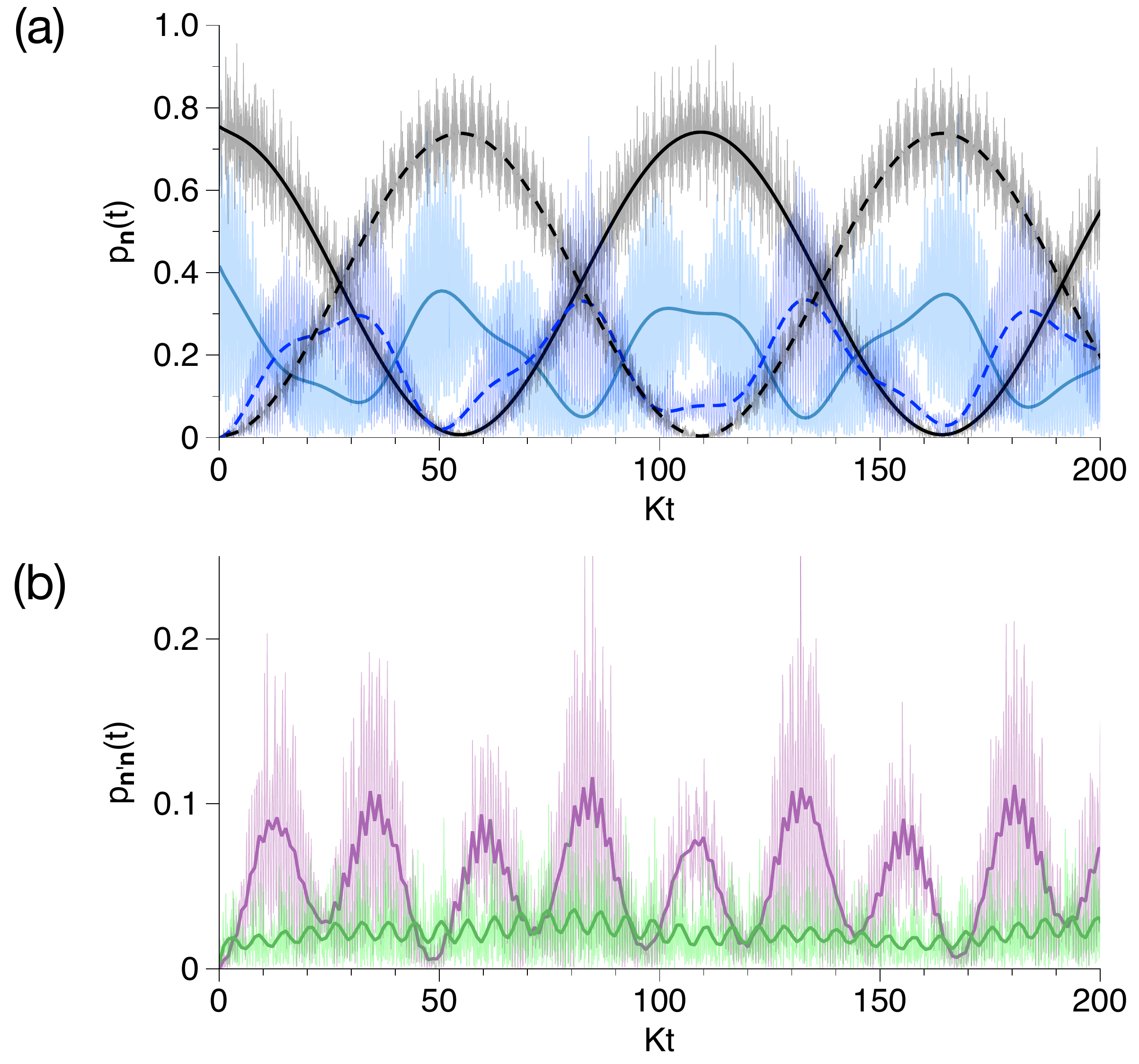} 
	\caption{(a) Survival probability for the Fock state $|15,0,10\rangle$ for the uncoupled ($K_c = 0$, black solid) and coupled ($K_c =0.05$, light blue solid) cases. The corresponding cross probabilities for the state $|10,0,15;40\rangle$ are shown as black dashed and dark blue dashed lines. Again, the fast oscillations occur on a timescale $Kt \sim 0.5$.
	(b) Cross probabilities for select states involved in the dynamics of the Fock state $|16,0,8;40\rangle$ in the coupled case with $K_c =0.05$.  The states shown are $|10,0,14;40\rangle$ (violet) and $|12,0,12;40\rangle$ (green). Note that, as before, all the smoothed probabilities in (a) and (b) are shown as bold lines.}
	\label{fig:15010prob}
\end{figure}

In Fig.~\ref{fig:15010prob}(a) we show the fragmentation dynamics of state $|15,0,10;40\rangle$. It is interesting to note that, in contrast to the case of $|16,0,8;40\rangle$, the survival probability  for the uncoupled trimer already shows coherent oscillations and  particle exchange between well one and three leading to a substantial probability flow to the state $|10,0,15;40\rangle$, a net five particle exchange. This particle exchange is reflected in Fig.~\ref{fig:imb} as $X_1$  shows coherent oscillation even for $K_{c}=0$. The timescale associated with this quantum transport is $Kt \sim 55$ and the  hopping term $(\hat{a}_1^{\dagger} \hat{a}_3 + h.c.)$ must be responsible. However, it is crucial to note that these two states are not connected classically for $K_{c}=0$, thus implicating the RAT mechanism. In contrast, for $K_{c} \neq 0$  multiple timescales emerge and the probability flows between the states $|15,0,10;40\rangle$ and $|10,0,15;40\rangle$ on a faster timescale $Kt \sim 16$. The presence of multiple timescales suggests the availability of other pathways due to involvement of the monomer. As example, we show in Fig.~\ref{fig:15010prob}(b) two of the states which participate in the dynamics. The fact that the population imbalance $X_1$ in Fig.~\ref{fig:imb} exhibits decay classically suggests that for $K_{c} \neq 0$ the states involved in the dynamics are connected by classical pathways as well. Our computations (not shown) indicate that classical mixing between the states exists, albeit dynamical tunneling still is the dominant mechanism.

\subsubsection{Tying it all together: Arnold web perspective}
We now argue that the results of Fig.~\ref{fig:imb} and the associated quantum probabilities seen in Figs.~\ref{fig:1706surv}-\ref{fig:15010prob} can be rationalized based on the relative locations of the three states on the Arnold web. It is useful to note that all three states of interest have $n_{0}=n_{1}$ and hence one expects the monomer-trimer resonance $R_{M1}$ plays a key role in the dynamics. Moreover, with decreasing $n_{1}$ the states approach the condition $n_{1} = n_{3}$ i.e., the trimer resonance $R_{T2}$. In fact, classically the following resonances (with constraint $I_{2}=0$) are expected to be important:
\begin{eqnarray}
R_{T2}&:& I_{0} = {\cal N} - 2I_{1} \\
R_{M1}&:& I_{0} = I_{1} \\
R_{M3}&:& 2I_{0} = {\cal N} - I_{1}
\end{eqnarray}
The first two resonances above intersect at $(I_{1}^{\times},I_{0}^{\times}) = ({\cal N}/3,{\cal N}/3)$ and the third can be considered as induced by the junction. 

\begin{figure*}
    \centering
    \includegraphics[width=1.0\linewidth]{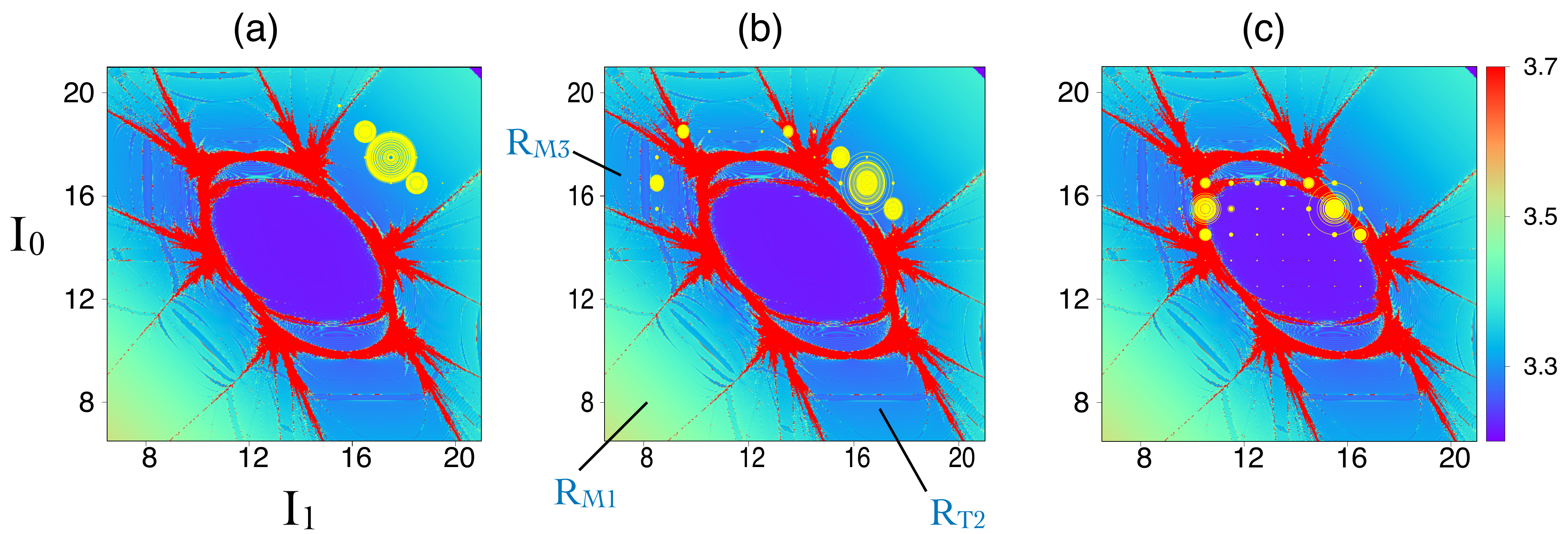}
    \caption{The Arnold web in the $(I_{1},I_{0})$ space corresponding to the Fock states (a) $|17,0,6;40\rangle$ (b) $|16,0,8;40\rangle$ and (c) $|15,0,10;40\rangle$. These webs have been constructed using the FLI measure on a $500 \times 500$ grid of initial conditions with the constraint $I_{2}=0$. The multiplicity-$2$ junction at $(I_{1},I_{0}) \approx (14,14)$ formed by the intersection of the resonances $R_{M1}, R_{T2}$, and $R_{M2}$ can be clearly seen. In every plot the quantum probability flow to different participating Fock states at times $Kt=0,10,20,\ldots,1000$ are shown as yellow circles, with radius $\propto$ probability.}
    \label{fig:figfockweb}
\end{figure*}

In Fig.~\ref{fig:figfockweb} the FLI-based computation of the Arnold web is shown projected on the $(I_{1},I_{0})$ space. In order to make contact with the quantum results discussed in the previous section, we also show the flow of the quantum probability in the action space superimposed on the Arnold web. Note that the Fock states being accessed at regular time intervals up to a final time $Kt=1000$ are shown as circles with radii proportional to the respective probabilities. Thus, one can readily get information about the dominant states and potential pathways that utilize specific resonances.

It is clear that the state $|17,0,6;40\rangle$ is in the $R_{M1}$ resonance and away from the junction. Since the $R_{M1}$ resonance channel is present only for $K_{c} \neq 0$, the Arnold web and dynamics in Fig.~\ref{fig:figfockweb}(a) explains the survival results in Fig.~\ref{fig:1706surv} - the states $|n_{1},0,6;40\rangle$ with $n_{1}=16,18$ are located within the $R_{M1}$ zone and hence accessible both classically and quantum mechanically. At the other end, the state $|15,0,10;40\rangle$ is under the influence of the junction i.e., multiple resonances. Consequently, the local chaos near the junction leads to classical mixing as well. Note that this also explains the survival results in Fig.~\ref{fig:15010prob}(a) for $K_{c}=0$. In this case the resonance $R_{M1}$  and $R_{M3}$ are absent and only the trimer resonance $R_{T2}$ is present. The location of the two states $|15,0,10;40\rangle$ and $|10,0,15;40\rangle$ is almost symmetric with respect to the $R_{T2}$ resonance center which leads to RAT connecting the states. It is hence not surprising that there is no classical pathway connecting the two states. With $K_{c} \neq 0$ the junction forms and the local chaos results in classical pathways opening up. Dynamical tunneling still occurs but becomes less coherent due to the competing classical pathways.

The Fock state $|16,0,8;40\rangle$ can be considered as dynamically being in the ``transition" zone - the coupling of the monomer   opens up RAT pathways that are unavailable to the state $|17,0,6;40\rangle$, but there is still no access to the classical pathways that are available to the state $|15,0,10;40\rangle$. Therefore, for $K_{c} \neq 0$, the state is protected classically but not quantum mechanically. Finally, the  observation made previously that the $|15,0,8;40\rangle \rightarrow |13,0,9;40\rangle$  classically forbidden transition requires a term like $\hat{a}_{1}^{2}\hat{a}_{3}^{\dagger}\hat{a}_{0}^{\dagger}$ is easy to understand. Observe that $\hat{a}_{1}^{2}\hat{a}_{3}^{\dagger}\hat{a}_{0}^{\dagger}\equiv \hat{a}_{1}\hat{a}_{3}^{\dagger}\hat{a}_{1}\hat{a}_{0}^{\dagger}$ corresponds to the fourth order resonance $2\Omega_{1}-\Omega_{3}-\Omega_{0} = 0$. Such a higher order resonance is precisely expected to be induced at the multiplicity-$2$ junction formed by the resonances $\Omega_{1}-\Omega_{3}=0$ ($R_{T2}$) and $\Omega_{1}-\Omega_{0}=0$ ($R_{M1}$). Indeed we have perturbatively confirmed that such a resonance is induced at the junction which gives rise to RAT between $|15,0,8;40\rangle$ and $|13,0,9;40\rangle$.

\section{Conclusions and outlook}
\label{sect:conc}

In this study we have analyzed a minimal bipartite (trimer plus monomer) Bose-Hubbard model, utilized earlier for highlighting the role of chaos in the thermalization of a many-body system, from a detailed classical-quantum perspective. For this purpose we have explicitly mapped the Arnold web of the system using the technique of fast Lyapunov indicator. We have demonstrated that the quantum dynamics senses the classical Arnold web. As a consequence the initial trimer configurations that are robust under the monomer perturbations can be predicted. Furthermore, due to the richness of the features on the Arnold web, various pathways open up for utilizing the weak monomer coupling to manipulate the transport within the trimer subsystem. For instance, the monomer can lead to the inversion of the trimer configuration. In addition, the timescale of transport within the trimer subsystem can be modulated by coupling to the monomer. Note that the results of Fig.~\ref{fig:figfockweb} provide the dynamical counterpart to the approximate phase space domains identified by Nigro et al in their study (see Fig.~9 in ref~\cite{Nigro_PRA_2018}). It would be interesting to study  the ``diffusion" on the Arnold web and compare to the Fokker-Planck diffusion equation model\cite{KhripkovJPCA2016} for the trimer population distribution.  In this context, based on the recent results\cite{ManikandanKS,KarmakarJPCA2018,SKPKYKS2020}, the role of the resonance junctions needs to be explored further since the effects of the junction can persist for stronger couplings as well.  Therefore we anticipate that further detailed investigation of the dynamics near the junction will yield valuable insights into the issue of thermalization. Although the  model studied is not the most general four-site system, the techniques and analysis presented in this work can be extended to other configurations. For instance, in a four-site ``plaquette" configuration the primary resonance $R_{M1}$ is absent. However, such a resonance can be induced (cf. Fig.~\ref{fig:figfockweb}) at a higher order due to the intersection of the $R_{T2}$ and the $R_{M3}$ resonances. Alternatively, other designed resonances can be induced by appropriate external driving fields. Thus, for example, it would be interesting to analyze the Arnold web features  of the four-site extended Bose-Hubbard model recently studied by Gr\"{u}n et al\cite{grun2021atomtronic} in the context of NOON state generation.

A further observation that arises from the current work is that the proximity of an initial state to a resonance junction  can trigger quite different fragmentation dynamics for trimer configurations which are very similar. As shown here, the differences arise due to the extent to which dynamical tunneling (resonance-assisted and potentially chaos-assisted) dominates the fragmentation process. Recently, Vanhaele and Schlagheck have shown\cite{VanhaelePRA2021} that robust NOON states can be generated efficiently by utlizing a combination of CAT and RAT mechanisms in a driven Bose-Hubbard dimer. Similarly, Arnal et al have experimentally demonstrated\cite{Arnaleabc4886} the role of CAT in designing tunable effective superlattices. Moreover, the dynamics of a periodically driven trimer\cite{Stiebler_2011} has a higher dimensional phase space and is best understood in terms of the Arnold web. In this regard our results in Fig.~\ref{fig:15010prob}(a) indicate that being in the vicinity of a resonance junction can lead to a fast ($Kt \sim 30$) exchange of bosons between the two trimer wells. Thus, engineering specific resonance junctions on the Arnold web by modulating the driving frequencies may allow for the creation of novel multimode NOON states. Another example that may benefit from the current approach is the study by Richaud and Penna\cite{Richaud_2018} on the phase separation dynamics of a bosonic binary mixture in a ring trimer model. It would be interesting to see if the features on the Arnold web for this four degrees of freedom system correlate with  the competition between phase separation and locally chaotic dynamics. We point out that a rich enough Arnold web structure emerges only when the number of particles in the BEC is sufficiently large i.e., in the semiclassical limit.  Nevertheless, semiclassical investigations of thermalization and dynamical tunneling in multi-site and multi-species Bose-Hubbard models from the Arnold web perspective may prove worthwhile.

\section{Acknowledgement}
S. Karmakar thanks the University Grants Commission (UGC), India for a doctoral fellowship. This work is supported by the Science and Engineering Research Board (SERB) of India under the grant EMR/2016/006246. We acknowledge the IIT Kanpur High Performance Computing facility for computing resources.

\appendix

\section{Bose-Hubbard Hamiltonian for BEC in an external trap}

In this appendix 
we provide  a brief derivation of a general Bose-Hubbard Hamiltonian (BHH). Details of the derivation
and validity of the various approximations
can be found in several works.
Consider the Hamiltonian for $N$ weakly interacting
spinless bosons in an external trapping potential $V_{ex}({\bf r})$
\begin{eqnarray}
H&=&H_{0} + H_{I} \nonumber \\
H_{0}&=&\int d{\bf r} \Psi^{\dagger}({\bf r}) 
\left[-\frac{\hbar^{2}}{2m} \nabla^{2} + V_{ex}({\bf r}) \right]
\Psi({\bf r}) \\
H_{I}&=& \frac{1}{2} \int \int d{\bf r} d{\bf r}' \Psi^{\dagger}({\bf r})
\Psi^{\dagger}({\bf r}') V({\bf r}-{\bf r}') \Psi({\bf r}') \Psi({\bf r})
\nonumber
\label{becham}
\end{eqnarray}
with $V({\bf r}-{\bf r}')$ being the two-body interaction term and
$\Psi^{\dagger}({\bf r}),\Psi({\bf r})$ being the field creation-annhilation
operators satisfying the commutation rule 
$[\Psi({\bf r}),\Psi^{\dagger}({\bf r}')]=\delta^{3}({\bf r}-{\bf r}')$.
Under cold and dilute conditions typically the two-body term is replaced
with a point-like approximation
\begin{equation}
V({\bf r}-{\bf r}') \approx \frac{4\pi a_{s} \hbar^{2}}{m}
\delta^{3}({\bf r}-{\bf r}') \equiv g \delta^{3}({\bf r}-{\bf r}')
\end{equation}
with $a_{s}$ being the $s$-wave scattering length. Thus, one obtains
a simplified interaction Hamiltonian $H_{I}$ 
\begin{equation}
H_{I}=\frac{1}{2} g \int d{\bf r} \Psi^{\dagger}({\bf r})
\Psi^{\dagger}({\bf r})\Psi({\bf r}) \Psi({\bf r})
\label{hamintappx}
\end{equation}
Furthermore a mean-field treatment is motivated by expressing the
field operator as
\begin{equation}
\Psi({\bf r}) = \langle \Psi({\bf r}) \rangle + \Psi'({\bf r})
\end{equation}
where, $\langle \Psi({\bf r}) \rangle \equiv \phi({\bf r})$ is
the order parameter (also called as the condensate wavefunction)
and $\Psi'({\bf r})$ accounts for the fluctuations leading to
the depletion of the BEC condensate. The classical field
$\phi({\bf r})$  yields the condensate density and obeys the
celebrated Gross-Pitaevskii equation (GPE). 

In this work, however, the focus is on the Bose-Hubbard description
rather than the GPE. The BHH mapping is obtained by expanding
the field operator in some convenient basis. For a BEC trapped
in a $n$-well (site) external potential
one can utilize a localized 
single particle basis ${\varphi_{n}({\bf r})}$ and write
\begin{equation}
\Psi({\bf r}) = \sum_{n;\alpha} \varphi_{n \alpha}({\bf r}) a_{n \alpha}
\end{equation}
with $\alpha$ indexing the states within a given well. 
The operator $a_{n \alpha}$ is the usual bosonic annhilation
operator satisfying the commutation relation 
$[a_{n \alpha},a^{\dagger}_{n' \alpha}]=\delta_{nn'}$. In the low temperature dilute gas limit, only the lowest state
in each well is relevant (the so called lowest band approximation) and thus the index $\alpha$ can be safely
ignored from our discussions. Substituting the above expansion into
the Hamiltonian in Eq.~\ref{becham}, with the form of $H_{I}$ shown
in Eq.~\ref{hamintappx}, one obtains
\begin{equation}
H = \sum_{mn} \epsilon_{nm} a^{\dagger}_{n}a_{m}
    + \frac{1}{2} g \sum_{klmn} T_{klmn} a^{\dagger}_{k} a^{\dagger}_{l}
                                         a_{m} a_{n}
\label{bhhgen}
\end{equation}
with
\begin{eqnarray}
\epsilon_{nm} &=& \int d{\bf r} \varphi_{n}^{*}({\bf r})
\left[-\frac{\hbar^{2}}{2m} \nabla^{2} + V_{ex}({\bf r}) \right]
\varphi_{m}({\bf r}) \\
T_{klmn} &=& \int d{\bf r} \varphi_{k}^{*}({\bf r}) \varphi_{l}^{*}({\bf r})
\varphi_{m}({\bf r}) \varphi_{n}({\bf r})
\end{eqnarray}
The BHH system in Eq.~\ref{bhhgen} arises in many different contexts,
characterized by different origins and interpretations of the
parameters $(\epsilon_{nm},T_{klmn})$.






\bibliographystyle{elsarticle-num-names}
\bibliography{references.bib}

\begin{thebibliography}{65}
\expandafter\ifx\csname natexlab\endcsname\relax\def\natexlab#1{#1}\fi
\providecommand{\url}[1]{\texttt{#1}}
\providecommand{\href}[2]{#2}
\providecommand{\path}[1]{#1}
\providecommand{\DOIprefix}{doi:}
\providecommand{\ArXivprefix}{arXiv:}
\providecommand{\URLprefix}{URL: }
\providecommand{\Pubmedprefix}{pmid:}
\providecommand{\doi}[1]{\href{http://dx.doi.org/#1}{\path{#1}}}
\providecommand{\Pubmed}[1]{\href{pmid:#1}{\path{#1}}}
\providecommand{\bibinfo}[2]{#2}
\ifx\xfnm\relax \def\xfnm[#1]{\unskip,\space#1}\fi
\bibitem[{Anderson et~al.(1995)Anderson, Ensher, Matthews, Wieman, and
  Cornell}]{Anderson198}
\bibinfo{author}{M.~H. Anderson}, \bibinfo{author}{J.~R. Ensher},
  \bibinfo{author}{M.~R. Matthews}, \bibinfo{author}{C.~E. Wieman},
  \bibinfo{author}{E.~A. Cornell},
\newblock \bibinfo{title}{Observation of \uppercase{B}ose-\uppercase{E}instein
  condensation in a dilute atomic vapor},
\newblock \bibinfo{journal}{Science} \bibinfo{volume}{269}
  (\bibinfo{year}{1995}) \bibinfo{pages}{198--201}.
\bibitem[{Davis et~al.(1995)Davis, Mewes, Andrews, van Druten, Durfee, Kurn,
  and Ketterle}]{KetterlePRL95}
\bibinfo{author}{K.~B. Davis}, \bibinfo{author}{M.~O. Mewes},
  \bibinfo{author}{M.~R. Andrews}, \bibinfo{author}{N.~J. van Druten},
  \bibinfo{author}{D.~S. Durfee}, \bibinfo{author}{D.~M. Kurn},
  \bibinfo{author}{W.~Ketterle},
\newblock \bibinfo{title}{Bose-\uppercase{E}instein condensation in a gas of
  sodium atoms},
\newblock \bibinfo{journal}{Phys. Rev. Lett.} \bibinfo{volume}{75}
  (\bibinfo{year}{1995}) \bibinfo{pages}{3969--3973}.
\bibitem[{Bradley et~al.(1995)Bradley, Sackett, Tollett, and
  Hulet}]{HuletPRL95}
\bibinfo{author}{C.~C. Bradley}, \bibinfo{author}{C.~A. Sackett},
  \bibinfo{author}{J.~J. Tollett}, \bibinfo{author}{R.~G. Hulet},
\newblock \bibinfo{title}{Evidence of \uppercase{B}ose-\uppercase{E}instein
  condensation in an atomic gas with attractive interactions},
\newblock \bibinfo{journal}{Phys. Rev. Lett.} \bibinfo{volume}{75}
  (\bibinfo{year}{1995}) \bibinfo{pages}{1687--1690}.
\bibitem[{Jaksch and Zoller(2005)}]{JAKSCH200552}
\bibinfo{author}{D.~Jaksch}, \bibinfo{author}{P.~Zoller},
\newblock \bibinfo{title}{The cold atom {H}ubbard toolbox},
\newblock \bibinfo{journal}{Annals of Physics} \bibinfo{volume}{315}
  (\bibinfo{year}{2005}) \bibinfo{pages}{52--79}. \bibinfo{note}{Special
  Issue}.
\bibitem[{Bloch(2005)}]{Bloch_NP_2005}
\bibinfo{author}{I.~Bloch},
\newblock \bibinfo{title}{Ultracold quantum gases in optical lattices},
\newblock \bibinfo{journal}{Nature Phys} \bibinfo{volume}{1}
  (\bibinfo{year}{2005}) \bibinfo{pages}{23--30}.
\bibitem[{Gati and Oberthaler(2007)}]{Gati_2007}
\bibinfo{author}{R.~Gati}, \bibinfo{author}{M.~K. Oberthaler},
\newblock \bibinfo{title}{A bosonic {J}osephson junction},
\newblock \bibinfo{journal}{Journal of Physics B: Atomic, Molecular and Optical
  Physics} \bibinfo{volume}{40} (\bibinfo{year}{2007})
  \bibinfo{pages}{R61--R89}.
\bibitem[{Zibold et~al.(2010)Zibold, Nicklas, Gross, and
  Oberthaler}]{ZiboldPRL2010}
\bibinfo{author}{T.~Zibold}, \bibinfo{author}{E.~Nicklas},
  \bibinfo{author}{C.~Gross}, \bibinfo{author}{M.~K. Oberthaler},
\newblock \bibinfo{title}{Classical bifurcation at the transition from
  \uppercase{R}abi to \uppercase{J}osephson dynamics},
\newblock \bibinfo{journal}{Phys. Rev. Lett.} \bibinfo{volume}{105}
  (\bibinfo{year}{2010}) \bibinfo{pages}{204101}.
\bibitem[{Aubry et~al.(1996)Aubry, Flach, Kladko, and Olbrich}]{OlbrichPRL1996}
\bibinfo{author}{S.~Aubry}, \bibinfo{author}{S.~Flach},
  \bibinfo{author}{K.~Kladko}, \bibinfo{author}{E.~Olbrich},
\newblock \bibinfo{title}{Manifestation of classical bifurcation in the
  spectrum of the integrable quantum dimer},
\newblock \bibinfo{journal}{Phys. Rev. Lett.} \bibinfo{volume}{76}
  (\bibinfo{year}{1996}) \bibinfo{pages}{1607--1610}.
\bibitem[{Kellman and Tyng(2002)}]{KellmanPRA2002}
\bibinfo{author}{M.~E. Kellman}, \bibinfo{author}{V.~Tyng},
\newblock \bibinfo{title}{Bifurcation effects in coupled
  \uppercase{B}ose-\uppercase{E}instein condensates},
\newblock \bibinfo{journal}{Phys. Rev. A} \bibinfo{volume}{66}
  (\bibinfo{year}{2002}) \bibinfo{pages}{013602}.
\bibitem[{Rubeni et~al.(2017)Rubeni, Links, Isaac, and
  Foerster}]{RubeniPRA2017}
\bibinfo{author}{D.~Rubeni}, \bibinfo{author}{J.~Links}, \bibinfo{author}{P.~S.
  Isaac}, \bibinfo{author}{A.~Foerster},
\newblock \bibinfo{title}{Two-site \uppercase{B}ose-\uppercase{H}ubbard model
  with nonlinear tunneling: Classical and quantum analysis},
\newblock \bibinfo{journal}{Phys. Rev. A} \bibinfo{volume}{95}
  (\bibinfo{year}{2017}) \bibinfo{pages}{043607}.
\bibitem[{Mahmud et~al.(2005)Mahmud, Perry, and Reinhardt}]{ReinhardtPRA2005}
\bibinfo{author}{K.~W. Mahmud}, \bibinfo{author}{H.~Perry},
  \bibinfo{author}{W.~P. Reinhardt},
\newblock \bibinfo{title}{Quantum phase-space picture of
  \uppercase{B}ose-\uppercase{E}instein condensates in a double well},
\newblock \bibinfo{journal}{Phys. Rev. A} \bibinfo{volume}{71}
  (\bibinfo{year}{2005}) \bibinfo{pages}{023615}.
\bibitem[{Chuchem et~al.(2010)Chuchem, Smith-Mannschott, Hiller, Kottos, Vardi,
  and Cohen}]{MayaCohenPRA2010}
\bibinfo{author}{M.~Chuchem}, \bibinfo{author}{K.~Smith-Mannschott},
  \bibinfo{author}{M.~Hiller}, \bibinfo{author}{T.~Kottos},
  \bibinfo{author}{A.~Vardi}, \bibinfo{author}{D.~Cohen},
\newblock \bibinfo{title}{Quantum dynamics in the bosonic \uppercase{J}osephson
  junction},
\newblock \bibinfo{journal}{Phys. Rev. A} \bibinfo{volume}{82}
  (\bibinfo{year}{2010}) \bibinfo{pages}{053617}.
\bibitem[{Nemoto et~al.(2000)Nemoto, Holmes, Milburn, and Munro}]{MunroPRA2000}
\bibinfo{author}{K.~Nemoto}, \bibinfo{author}{C.~A. Holmes},
  \bibinfo{author}{G.~J. Milburn}, \bibinfo{author}{W.~J. Munro},
\newblock \bibinfo{title}{Quantum dynamics of three coupled atomic
  \uppercase{B}ose-\uppercase{E}instein condensates},
\newblock \bibinfo{journal}{Phys. Rev. A} \bibinfo{volume}{63}
  (\bibinfo{year}{2000}) \bibinfo{pages}{013604}.
\bibitem[{Franzosi and Penna(2003)}]{PennaPRE2003}
\bibinfo{author}{R.~Franzosi}, \bibinfo{author}{V.~Penna},
\newblock \bibinfo{title}{Chaotic behavior, collective modes, and self-trapping
  in the dynamics of three coupled \uppercase{B}ose-\uppercase{E}instein
  condensates},
\newblock \bibinfo{journal}{Phys. Rev. E} \bibinfo{volume}{67}
  (\bibinfo{year}{2003}) \bibinfo{pages}{046227}.
\bibitem[{Liu et~al.(2007)Liu, Fu, Yang, and Liu}]{LiuPRA2007}
\bibinfo{author}{B.~Liu}, \bibinfo{author}{L.-B. Fu}, \bibinfo{author}{S.-P.
  Yang}, \bibinfo{author}{J.~Liu},
\newblock \bibinfo{title}{Josephson oscillation and transition to self-trapping
  for \uppercase{B}ose-\uppercase{E}instein condensates in a triple-well trap},
\newblock \bibinfo{journal}{Phys. Rev. A} \bibinfo{volume}{75}
  (\bibinfo{year}{2007}) \bibinfo{pages}{033601}.
\bibitem[{Dey et~al.(2018)Dey, Cohen, and Vardi}]{DeyVardiPRL2018}
\bibinfo{author}{A.~Dey}, \bibinfo{author}{D.~Cohen},
  \bibinfo{author}{A.~Vardi},
\newblock \bibinfo{title}{Adiabatic passage through chaos},
\newblock \bibinfo{journal}{Phys. Rev. Lett.} \bibinfo{volume}{121}
  (\bibinfo{year}{2018}) \bibinfo{pages}{250405}.
\bibitem[{Smerzi et~al.(1997)Smerzi, Fantoni, Giovanazzi, and
  Shenoy}]{SmerziPRL1997}
\bibinfo{author}{A.~Smerzi}, \bibinfo{author}{S.~Fantoni},
  \bibinfo{author}{S.~Giovanazzi}, \bibinfo{author}{S.~R. Shenoy},
\newblock \bibinfo{title}{Quantum coherent atomic tunneling between two trapped
  {B}ose-{E}instein condensates},
\newblock \bibinfo{journal}{Phys. Rev. Lett.} \bibinfo{volume}{79}
  (\bibinfo{year}{1997}) \bibinfo{pages}{4950--4953}.
\bibitem[{Albiez et~al.(2005)Albiez, Gati, F\"olling, Hunsmann, Cristiani, and
  Oberthaler}]{Oberthaler2005}
\bibinfo{author}{M.~Albiez}, \bibinfo{author}{R.~Gati},
  \bibinfo{author}{J.~F\"olling}, \bibinfo{author}{S.~Hunsmann},
  \bibinfo{author}{M.~Cristiani}, \bibinfo{author}{M.~K. Oberthaler},
\newblock \bibinfo{title}{Direct observation of tunneling and nonlinear
  self-trapping in a single bosonic {J}osephson junction},
\newblock \bibinfo{journal}{Phys. Rev. Lett.} \bibinfo{volume}{95}
  (\bibinfo{year}{2005}) \bibinfo{pages}{010402}.
\bibitem[{Raghavan et~al.(1999)Raghavan, Smerzi, Fantoni, and
  Shenoy}]{RaghavanPRA1999}
\bibinfo{author}{S.~Raghavan}, \bibinfo{author}{A.~Smerzi},
  \bibinfo{author}{S.~Fantoni}, \bibinfo{author}{S.~R. Shenoy},
\newblock \bibinfo{title}{Coherent oscillations between two weakly coupled
  {B}ose-{E}instein condensates: {J}osephson effects, $\ensuremath{\pi}$
  oscillations, and macroscopic quantum self-trapping},
\newblock \bibinfo{journal}{Phys. Rev. A} \bibinfo{volume}{59}
  (\bibinfo{year}{1999}) \bibinfo{pages}{620--633}.
\bibitem[{Kolovsky(2007)}]{kolovsky2007}
\bibinfo{author}{A.~R. Kolovsky},
\newblock \bibinfo{title}{Semiclassical quantization of the {B}ogoliubov
  spectrum},
\newblock \bibinfo{journal}{Phys. Rev. Lett.} \bibinfo{volume}{99}
  (\bibinfo{year}{2007}) \bibinfo{pages}{020401}.
\bibitem[{Stickney et~al.(2007)Stickney, Anderson, and
  Zozulya}]{Stickney_PRA2007}
\bibinfo{author}{J.~A. Stickney}, \bibinfo{author}{D.~Z. Anderson},
  \bibinfo{author}{A.~A. Zozulya},
\newblock \bibinfo{title}{Transistor like behavior of a {B}ose-{E}instein
  condensate in a triple-well potential},
\newblock \bibinfo{journal}{Phys. Rev. A} \bibinfo{volume}{75}
  (\bibinfo{year}{2007}) \bibinfo{pages}{013608}.
\bibitem[{Wilsmann et~al.(2018)Wilsmann, Ymai, Tonel, Links, and
  Foerster}]{Wilsmann_CommPhys2018}
\bibinfo{author}{K.~W. Wilsmann}, \bibinfo{author}{L.~H. Ymai},
  \bibinfo{author}{A.~P. Tonel}, \bibinfo{author}{J.~Links},
  \bibinfo{author}{A.~Foerster},
\newblock \bibinfo{title}{Control of tunneling in an atomtronic switching
  device},
\newblock \bibinfo{journal}{Commun Phys} \bibinfo{volume}{1}
  (\bibinfo{year}{2018}) \bibinfo{pages}{91}.
\bibitem[{Schlagheck et~al.(2010)Schlagheck, Malet, Cremon, and
  Reimann}]{Schlagheck_2010}
\bibinfo{author}{P.~Schlagheck}, \bibinfo{author}{F.~Malet},
  \bibinfo{author}{J.~C. Cremon}, \bibinfo{author}{S.~M. Reimann},
\newblock \bibinfo{title}{Transport and interaction blockade of cold bosonic
  atoms in a triple-well potential},
\newblock \bibinfo{journal}{New Journal of Physics} \bibinfo{volume}{12}
  (\bibinfo{year}{2010}) \bibinfo{pages}{065020}.
\bibitem[{Chianca and Olsen(2011)}]{OlsenPRA2011}
\bibinfo{author}{C.~V. Chianca}, \bibinfo{author}{M.~K. Olsen},
\newblock \bibinfo{title}{Quantum dynamics of a four-well
  \uppercase{B}ose-\uppercase{H}ubbard model with two different tunneling
  rates},
\newblock \bibinfo{journal}{Phys. Rev. A} \bibinfo{volume}{83}
  (\bibinfo{year}{2011}) \bibinfo{pages}{043607}.
\bibitem[{Khripkov and Vardi(2014)}]{KhripkovPRA2014}
\bibinfo{author}{C.~Khripkov}, \bibinfo{author}{A.~Vardi},
\newblock \bibinfo{title}{Coherence oscillations between weakly coupled
  \uppercase{B}ose-\uppercase{H}ubbard dimers},
\newblock \bibinfo{journal}{Phys. Rev. A} \bibinfo{volume}{89}
  (\bibinfo{year}{2014}) \bibinfo{pages}{053629}.
\bibitem[{Khripkov et~al.(2016)Khripkov, Cohen, and Vardi}]{KhripkovJPCA2016}
\bibinfo{author}{C.~Khripkov}, \bibinfo{author}{D.~Cohen},
  \bibinfo{author}{A.~Vardi},
\newblock \bibinfo{title}{Thermalization of bipartite
  \uppercase{B}ose–\uppercase{H}ubbard models},
\newblock \bibinfo{journal}{J. Phys. Chem. A} \bibinfo{volume}{120}
  (\bibinfo{year}{2016}) \bibinfo{pages}{3136--3141}.
\bibitem[{Khripkov et~al.(2018)Khripkov, Vardi, and Cohen}]{KhripkovPRE2018}
\bibinfo{author}{C.~Khripkov}, \bibinfo{author}{A.~Vardi},
  \bibinfo{author}{D.~Cohen},
\newblock \bibinfo{title}{Semiclassical theory of strong localization for
  quantum thermalization},
\newblock \bibinfo{journal}{Phys. Rev. E} \bibinfo{volume}{97}
  (\bibinfo{year}{2018}) \bibinfo{pages}{022127}.
\bibitem[{Khripkov et~al.(2015)Khripkov, Vardi, and Cohen}]{KhripkovNJP2015}
\bibinfo{author}{C.~Khripkov}, \bibinfo{author}{A.~Vardi},
  \bibinfo{author}{D.~Cohen},
\newblock \bibinfo{title}{Quantum thermalization: anomalous slow relaxation due
  to percolation-like dynamics},
\newblock \bibinfo{journal}{New J. Phys.} \bibinfo{volume}{17}
  (\bibinfo{year}{2015}) \bibinfo{pages}{023071}.
\bibitem[{Wiggins(1992)}]{wigbook}
\bibinfo{author}{S.~Wiggins}, \bibinfo{title}{Chaotic transport in dynamical
  systems}, \bibinfo{publisher}{Springer-Verlag: New York},
  \bibinfo{year}{1992}.
\bibitem[{Cincotta et~al.(2014)Cincotta, Efthymiopoulos, Giordano, and
  Mestre}]{Cincottaetal2014}
\bibinfo{author}{P.~M. Cincotta}, \bibinfo{author}{C.~Efthymiopoulos},
  \bibinfo{author}{C.~M. Giordano}, \bibinfo{author}{M.~F. Mestre},
\newblock \bibinfo{title}{{C}hirikov and {N}ekhoroshev diffusion estimates:
  Bridging the two sides of the river},
\newblock \bibinfo{journal}{Physica D} \bibinfo{volume}{266}
  (\bibinfo{year}{2014}) \bibinfo{pages}{49--64}.
\bibitem[{Efthymiopoulos and Harsoula(2013)}]{EFTHYMIOPOULOS201319}
\bibinfo{author}{C.~Efthymiopoulos}, \bibinfo{author}{M.~Harsoula},
\newblock \bibinfo{title}{The speed of {A}rnold diffusion},
\newblock \bibinfo{journal}{Physica D} \bibinfo{volume}{251}
  (\bibinfo{year}{2013}) \bibinfo{pages}{19--38}.
\bibitem[{Brodier et~al.(2001)Brodier, Schlagheck, and
  Ullmo}]{brodieretalprl01}
\bibinfo{author}{O.~Brodier}, \bibinfo{author}{P.~Schlagheck},
  \bibinfo{author}{D.~Ullmo},
\newblock \bibinfo{title}{Resonance-assisted tunneling in near-integrable
  systems},
\newblock \bibinfo{journal}{Phys. Rev. Lett.} \bibinfo{volume}{87}
  (\bibinfo{year}{2001}) \bibinfo{pages}{064101}.
\bibitem[{Tomsovic and Ullmo(1994)}]{tomsovic94}
\bibinfo{author}{S.~Tomsovic}, \bibinfo{author}{D.~Ullmo},
\newblock \bibinfo{title}{Chaos-assisted tunneling},
\newblock \bibinfo{journal}{Phys. Rev. E} \bibinfo{volume}{50}
  (\bibinfo{year}{1994}) \bibinfo{pages}{145--162}.
\bibitem[{Karmakar and Keshavamurthy(2018)}]{KarmakarJPCA2018}
\bibinfo{author}{S.~Karmakar}, \bibinfo{author}{S.~Keshavamurthy},
\newblock \bibinfo{title}{Relevance of the resonance junctions on the {A}rnold
  web to dynamical tunneling and eigenstate delocalization},
\newblock \bibinfo{journal}{The Journal of Physical Chemistry A}
  \bibinfo{volume}{122} (\bibinfo{year}{2018}) \bibinfo{pages}{8636--8649}.
\bibitem[{Firmbach et~al.(2019)Firmbach, Fritzsch, Ketzmerick, and
  B\"acker}]{Firmbach_PRE_2019}
\bibinfo{author}{M.~Firmbach}, \bibinfo{author}{F.~Fritzsch},
  \bibinfo{author}{R.~Ketzmerick}, \bibinfo{author}{A.~B\"acker},
\newblock \bibinfo{title}{Resonance-assisted tunneling in four-dimensional
  normal-form {H}amiltonians},
\newblock \bibinfo{journal}{Phys. Rev. E} \bibinfo{volume}{99}
  (\bibinfo{year}{2019}) \bibinfo{pages}{042213}.
\bibitem[{Pittman et~al.(2016)Pittman, Tannenbaum, and Heller}]{PittmanJCP2016}
\bibinfo{author}{S.~M. Pittman}, \bibinfo{author}{E.~Tannenbaum},
  \bibinfo{author}{E.~J. Heller},
\newblock \bibinfo{title}{Dynamical tunneling versus fast diffusion for a
  non-convex {H}amiltonian},
\newblock \bibinfo{journal}{The Journal of Chemical Physics}
  \bibinfo{volume}{145} (\bibinfo{year}{2016}) \bibinfo{pages}{054303}.
\bibitem[{Manikandan and Keshavamurthy(2014)}]{ManikandanKS}
\bibinfo{author}{P.~Manikandan}, \bibinfo{author}{S.~Keshavamurthy},
\newblock \bibinfo{title}{Dynamical traps lead to the slowing down of
  intramolecular vibrational energy flow},
\newblock \bibinfo{journal}{Proc. Natl. Acad. Sci. U.S.A.}
  \bibinfo{volume}{111} (\bibinfo{year}{2014}) \bibinfo{pages}{14354--14359}.
\bibitem[{Karmakar et~al.(2020)Karmakar, Yadav, and
  Keshavamurthy}]{SKPKYKS2020}
\bibinfo{author}{S.~Karmakar}, \bibinfo{author}{P.~K. Yadav},
  \bibinfo{author}{S.~Keshavamurthy},
\newblock \bibinfo{title}{Stable chaos and delayed onset of statisticality in
  unimolecular dissociation reactions},
\newblock \bibinfo{journal}{Commun Chem} \bibinfo{volume}{3}
  (\bibinfo{year}{2020}) \bibinfo{pages}{4}.
\bibitem[{Karmakar and Keshavamurthy(2020)}]{Karmakar_pccp_perspective}
\bibinfo{author}{S.~Karmakar}, \bibinfo{author}{S.~Keshavamurthy},
\newblock \bibinfo{title}{Intramolecular vibrational energy redistribution and
  the quantum ergodicity transition: a phase space perspective},
\newblock \bibinfo{journal}{Phys. Chem. Chem. Phys.} \bibinfo{volume}{22}
  (\bibinfo{year}{2020}) \bibinfo{pages}{11139--11173}.
\bibitem[{Kellman and Tyng(2007)}]{Kellman2007}
\bibinfo{author}{M.~E. Kellman}, \bibinfo{author}{V.~Tyng},
\newblock \bibinfo{title}{The dance of molecules: New dynamical perspectives on
  highly excited molecular vibrations},
\newblock \bibinfo{journal}{Acc. Chem. Res.} \bibinfo{volume}{40}
  (\bibinfo{year}{2007}) \bibinfo{pages}{243--250}.
\bibitem[{Farantos et~al.(2009)Farantos, Schinke, Guo, and
  Joyeux}]{Farantos2009}
\bibinfo{author}{S.~C. Farantos}, \bibinfo{author}{R.~Schinke},
  \bibinfo{author}{H.~Guo}, \bibinfo{author}{M.~Joyeux},
\newblock \bibinfo{title}{Energy localization in molecules, bifurcation
  phenomena, and their spectroscopic signatures: The global view},
\newblock \bibinfo{journal}{Chem. Rev.} \bibinfo{volume}{109}
  (\bibinfo{year}{2009}) \bibinfo{pages}{4248--4271}.
\bibitem[{Manikandan et~al.(2009)Manikandan, Semparithi, and
  Keshavamurthy}]{manikandan2009}
\bibinfo{author}{P.~Manikandan}, \bibinfo{author}{A.~Semparithi},
  \bibinfo{author}{S.~Keshavamurthy},
\newblock \bibinfo{title}{Decoding the dynamical information embedded in highly
  excited vibrational eigenstates: State space and phase space viewpoints},
\newblock \bibinfo{journal}{J. Phys. Chem. A} \bibinfo{volume}{113}
  (\bibinfo{year}{2009}) \bibinfo{pages}{1717--1730}.
\bibitem[{Gruebele and Wolynes(2004)}]{MGPGWACR2004}
\bibinfo{author}{M.~Gruebele}, \bibinfo{author}{P.~G. Wolynes},
\newblock \bibinfo{title}{Vibrational energy flow and chemical reactions},
\newblock \bibinfo{journal}{Accounts of Chemical Research} \bibinfo{volume}{37}
  (\bibinfo{year}{2004}) \bibinfo{pages}{261--267}.
\bibitem[{Keshavamurthy(2013)}]{srihariIVR2013}
\bibinfo{author}{S.~Keshavamurthy},
\newblock \bibinfo{title}{Scaling perspective on intramolecular vibrational
  energy flow: Analogies, insights, and challenges},
\newblock \bibinfo{journal}{Adv. Chem. Phys.} \bibinfo{volume}{153}
  (\bibinfo{year}{2013}) \bibinfo{pages}{43--110}.
\bibitem[{Leitner(2015)}]{Leitner2015}
\bibinfo{author}{D.~M. Leitner},
\newblock \bibinfo{title}{Quantum ergodicity and energy flow in molecules},
\newblock \bibinfo{journal}{Adv. Phys.} \bibinfo{volume}{64}
  (\bibinfo{year}{2015}) \bibinfo{pages}{445--517}.
\bibitem[{Froeschl{\'e} et~al.(1997)Froeschl{\'e}, Lega, and
  Gonczi}]{Froeschle1997}
\bibinfo{author}{C.~Froeschl{\'e}}, \bibinfo{author}{E.~Lega},
  \bibinfo{author}{R.~Gonczi},
\newblock \bibinfo{title}{Fast \uppercase{L}yapunov indicators. application to
  asteroidal motion},
\newblock \bibinfo{journal}{Celest. Mech. Dynamical Astron.}
  \bibinfo{volume}{67} (\bibinfo{year}{1997}) \bibinfo{pages}{41--62}.
\bibitem[{Froeschl{\'e} et~al.(2000)Froeschl{\'e}, Guzzo, and
  Lega}]{FroeschleScience2000}
\bibinfo{author}{C.~Froeschl{\'e}}, \bibinfo{author}{M.~Guzzo},
  \bibinfo{author}{E.~Lega},
\newblock \bibinfo{title}{Graphical evolution of the \uppercase{A}rnold web:
  From order to chaos},
\newblock \bibinfo{journal}{Science} \bibinfo{volume}{289}
  (\bibinfo{year}{2000}) \bibinfo{pages}{2108--2110}.
\bibitem[{Skokos et~al.(2016)Skokos, Gottwald, and Laskar}]{chaosdetect}
\bibinfo{author}{C.~Skokos}, \bibinfo{author}{G.~A. Gottwald},
  \bibinfo{author}{J.~Laskar}, \bibinfo{title}{Chaos Detection and
  Predictability, Vol. 915 of Lecture Notes in Physics},
  \bibinfo{publisher}{Springer, Berlin, Heidelberg}, \bibinfo{year}{2016}.
\bibitem[{Skokos et~al.(2007)Skokos, Bountis, and Antonopoulos}]{SKOKOS200730}
\bibinfo{author}{C.~Skokos}, \bibinfo{author}{T.~Bountis},
  \bibinfo{author}{C.~Antonopoulos},
\newblock \bibinfo{title}{Geometrical properties of local dynamics in
  {H}amiltonian systems: The generalized alignment index ({GALI}) method},
\newblock \bibinfo{journal}{Physica D: Nonlinear Phenomena}
  \bibinfo{volume}{231} (\bibinfo{year}{2007}) \bibinfo{pages}{30--54}.
\bibitem[{Skokos et~al.(2004)Skokos, Antonopoulos, Bountis, and
  Vrahatis}]{Skokos_2004}
\bibinfo{author}{C.~Skokos}, \bibinfo{author}{C.~Antonopoulos},
  \bibinfo{author}{T.~C. Bountis}, \bibinfo{author}{M.~N. Vrahatis},
\newblock \bibinfo{title}{Detecting order and chaos in hamiltonian systems by
  the {SALI} method},
\newblock \bibinfo{journal}{Journal of Physics A: Mathematical and General}
  \bibinfo{volume}{37} (\bibinfo{year}{2004}) \bibinfo{pages}{6269--6284}.
\bibitem[{Barrio(2005)}]{BARRIO2005711}
\bibinfo{author}{R.~Barrio},
\newblock \bibinfo{title}{Sensitivity tools vs. {P}oincar\'{e} sections},
\newblock \bibinfo{journal}{Chaos, Solitons \& Fractals} \bibinfo{volume}{25}
  (\bibinfo{year}{2005}) \bibinfo{pages}{711--726}.
\bibitem[{Cincotta et~al.(2003)Cincotta, Giordano, and
  Sim\'{o}}]{CINCOTTA2003151}
\bibinfo{author}{P.~Cincotta}, \bibinfo{author}{C.~Giordano},
  \bibinfo{author}{C.~Sim\'{o}},
\newblock \bibinfo{title}{Phase space structure of multi-dimensional systems by
  means of the mean exponential growth factor of nearby orbits},
\newblock \bibinfo{journal}{Physica D: Nonlinear Phenomena}
  \bibinfo{volume}{182} (\bibinfo{year}{2003}) \bibinfo{pages}{151--178}.
\bibitem[{Cincotta et~al.(2021)Cincotta, Giordano, {Alves Silva}, and
  Beaug\'{e}}]{CINCOTTA2021132816}
\bibinfo{author}{P.~M. Cincotta}, \bibinfo{author}{C.~M. Giordano},
  \bibinfo{author}{R.~{Alves Silva}}, \bibinfo{author}{C.~Beaug\'{e}},
\newblock \bibinfo{title}{The {S}hannon entropy: An efficient indicator of
  dynamical stability},
\newblock \bibinfo{journal}{Physica D: Nonlinear Phenomena}
  \bibinfo{volume}{417} (\bibinfo{year}{2021}) \bibinfo{pages}{132816}.
\bibitem[{Barrio et~al.(2009)Barrio, Borczyk, and Breiter}]{BARRIO20091697}
\bibinfo{author}{R.~Barrio}, \bibinfo{author}{W.~Borczyk},
  \bibinfo{author}{S.~Breiter},
\newblock \bibinfo{title}{Spurious structures in chaos indicators maps},
\newblock \bibinfo{journal}{Chaos, Solitons \& Fractals} \bibinfo{volume}{40}
  (\bibinfo{year}{2009}) \bibinfo{pages}{1697--1714}.
\bibitem[{Bohigas et~al.(1993)Bohigas, Tomsovic, and Ullmo}]{bohigas93}
\bibinfo{author}{O.~Bohigas}, \bibinfo{author}{S.~Tomsovic},
  \bibinfo{author}{D.~Ullmo},
\newblock \bibinfo{title}{Manifestations of classical phase space structures in
  quantum mechanics},
\newblock \bibinfo{journal}{Phys. Rep.} \bibinfo{volume}{223}
  (\bibinfo{year}{1993}) \bibinfo{pages}{43 -- 133}.
\bibitem[{Eltschka and Schlagheck(2005)}]{eltschka05}
\bibinfo{author}{C.~Eltschka}, \bibinfo{author}{P.~Schlagheck},
\newblock \bibinfo{title}{Resonance- and chaos-assisted tunneling in mixed
  regular-chaotic systems},
\newblock \bibinfo{journal}{Phys. Rev. Lett.} \bibinfo{volume}{94}
  (\bibinfo{year}{2005}) \bibinfo{pages}{014101}.
\bibitem[{Keshavamurthy(2005{\natexlab{a}})}]{ksjcp05}
\bibinfo{author}{S.~Keshavamurthy},
\newblock \bibinfo{title}{On dynamical tunneling and classical resonances},
\newblock \bibinfo{journal}{J. Chem. Phys.} \bibinfo{volume}{122}
  (\bibinfo{year}{2005}{\natexlab{a}}) \bibinfo{pages}{114109}.
\bibitem[{Keshavamurthy(2005{\natexlab{b}})}]{KSPRE2005}
\bibinfo{author}{S.~Keshavamurthy},
\newblock \bibinfo{title}{Resonance-assisted tunneling in three degrees of
  freedom without discrete symmetry},
\newblock \bibinfo{journal}{Phys. Rev. E} \bibinfo{volume}{72}
  (\bibinfo{year}{2005}{\natexlab{b}}) \bibinfo{pages}{045203}.
\bibitem[{Nigro et~al.(2018)Nigro, Capuzzi, Cataldo, and
  Jezek}]{Nigro_PRA_2018}
\bibinfo{author}{M.~Nigro}, \bibinfo{author}{P.~Capuzzi},
  \bibinfo{author}{H.~M. Cataldo}, \bibinfo{author}{D.~M. Jezek},
\newblock \bibinfo{title}{Dynamics in multiple-well {B}ose-{E}instein
  condensates},
\newblock \bibinfo{journal}{Phys. Rev. A} \bibinfo{volume}{97}
  (\bibinfo{year}{2018}) \bibinfo{pages}{013626}.
\bibitem[{Keshavamurthy(2007)}]{Srihari2007}
\bibinfo{author}{S.~Keshavamurthy},
\newblock \bibinfo{title}{Dynamical tunnelling in molecules: quantum routes to
  energy flow},
\newblock \bibinfo{journal}{Int. Rev. Phys. Chem.} \bibinfo{volume}{26}
  (\bibinfo{year}{2007}) \bibinfo{pages}{521--584}.
\bibitem[{Grun et~al.(2021)Grun, Wittmann, Ymai, Links, and
  Foerster}]{grun2021atomtronic}
\bibinfo{author}{D.~S. Grun}, \bibinfo{author}{K.~W. Wittmann},
  \bibinfo{author}{L.~H. Ymai}, \bibinfo{author}{J.~Links},
  \bibinfo{author}{A.~Foerster}, \bibinfo{title}{Atomtronic protocol designs
  for {NOON} states}, \bibinfo{year}{2021}.
  \href{http://arxiv.org/abs/2102.02944}{{\tt arXiv:2102.02944}}.
\bibitem[{Vanhaele and Schlagheck(2021)}]{VanhaelePRA2021}
\bibinfo{author}{G.~Vanhaele}, \bibinfo{author}{P.~Schlagheck},
\newblock \bibinfo{title}{{NOON} states with ultracold bosonic atoms via
  resonance- and chaos-assisted tunneling},
\newblock \bibinfo{journal}{Phys. Rev. A} \bibinfo{volume}{103}
  (\bibinfo{year}{2021}) \bibinfo{pages}{013315}.
\bibitem[{Arnal et~al.(2020)Arnal, Chatelain, Martinez, Dupont, Giraud, Ullmo,
  Georgeot, Lemari{\'e}, Billy, and Gu{\'e}ry-Odelin}]{Arnaleabc4886}
\bibinfo{author}{M.~Arnal}, \bibinfo{author}{G.~Chatelain},
  \bibinfo{author}{M.~Martinez}, \bibinfo{author}{N.~Dupont},
  \bibinfo{author}{O.~Giraud}, \bibinfo{author}{D.~Ullmo},
  \bibinfo{author}{B.~Georgeot}, \bibinfo{author}{G.~Lemari{\'e}},
  \bibinfo{author}{J.~Billy}, \bibinfo{author}{D.~Gu{\'e}ry-Odelin},
\newblock \bibinfo{title}{Chaos-assisted tunneling resonances in a synthetic
  {F}loquet superlattice},
\newblock \bibinfo{journal}{Science Advances} \bibinfo{volume}{6}
  (\bibinfo{year}{2020}).
\bibitem[{Stiebler et~al.(2011)Stiebler, Gertjerenken, Teichmann, and
  Weiss}]{Stiebler_2011}
\bibinfo{author}{K.~Stiebler}, \bibinfo{author}{B.~Gertjerenken},
  \bibinfo{author}{N.~Teichmann}, \bibinfo{author}{C.~Weiss},
\newblock \bibinfo{title}{Spatial two-particle {NOON}-states in periodically
  shaken three-well potentials},
\newblock \bibinfo{journal}{Journal of Physics B: Atomic, Molecular and Optical
  Physics} \bibinfo{volume}{44} (\bibinfo{year}{2011}) \bibinfo{pages}{055301}.
\bibitem[{Richaud and Penna(2018)}]{Richaud_2018}
\bibinfo{author}{A.~Richaud}, \bibinfo{author}{V.~Penna},
\newblock \bibinfo{title}{Phase separation can be stronger than chaos},
\newblock \bibinfo{journal}{New Journal of Physics} \bibinfo{volume}{20}
  (\bibinfo{year}{2018}) \bibinfo{pages}{105008}.

\end{thebibliography}







\end{document}